\documentclass[12pt]{article}
\usepackage{graphicx}
\usepackage{psfrag}
\usepackage{amsmath}
\usepackage{amssymb}
\usepackage{amsthm}

\newtheorem{conjecture}{Conjecture}

\newtheorem{definition}{Definition}

\newtheorem{theorem}{Theorem}

\newtheorem{proposition}{Proposition}
\newtheorem{example}{Example}

\begin{document}

\title{Connecting period-doubling cascades to chaos}

\author{Evelyn Sander and James A. Yorke}

\maketitle

\begin{abstract}
  The appearance of infinitely-many period-doubling cascades is one
  of the most prominent features observed in the study of maps
  depending on a parameter. They are associated with chaotic behavior,
  since bifurcation diagrams of a map with a parameter often reveal a
  complicated intermingling of period-doubling cascades and chaos.

  Period doubling can be studied at three levels of complexity. The
  first is an individual period-doubling bifurcation. The second is an
  infinite collection of period doublings that are connected together
  by periodic orbits in a pattern called a cascade. It was first
  described by Myrberg and later in more detail by Feigenbaum. The
  third involves infinitely many cascades and a parameter value
  $\mu_2$ of the map at which there is chaos. We show that often
  virtually all (i.e., all but finitely many) ``regular''
  periodic orbits at $\mu_2$ are each connected to exactly one cascade
  by a path of regular periodic orbits; and virtually all cascades are
  either paired -- connected to exactly one other cascade, or solitary
  -- connected to exactly one regular periodic orbit at $\mu_2$. The
  solitary cascades are robust to large perturbations.  Hence the
  investigation of infinitely many cascades is
  essentially reduced to studying the regular periodic orbits of
  $F(\mu_2, \cdot)$. Examples discussed include the forced-damped
  pendulum and the double-well Duffing equation.

\end{abstract}

\section{Introduction}

\begin{figure}[t]
\begin{center}
\includegraphics[width=1.0\textwidth]{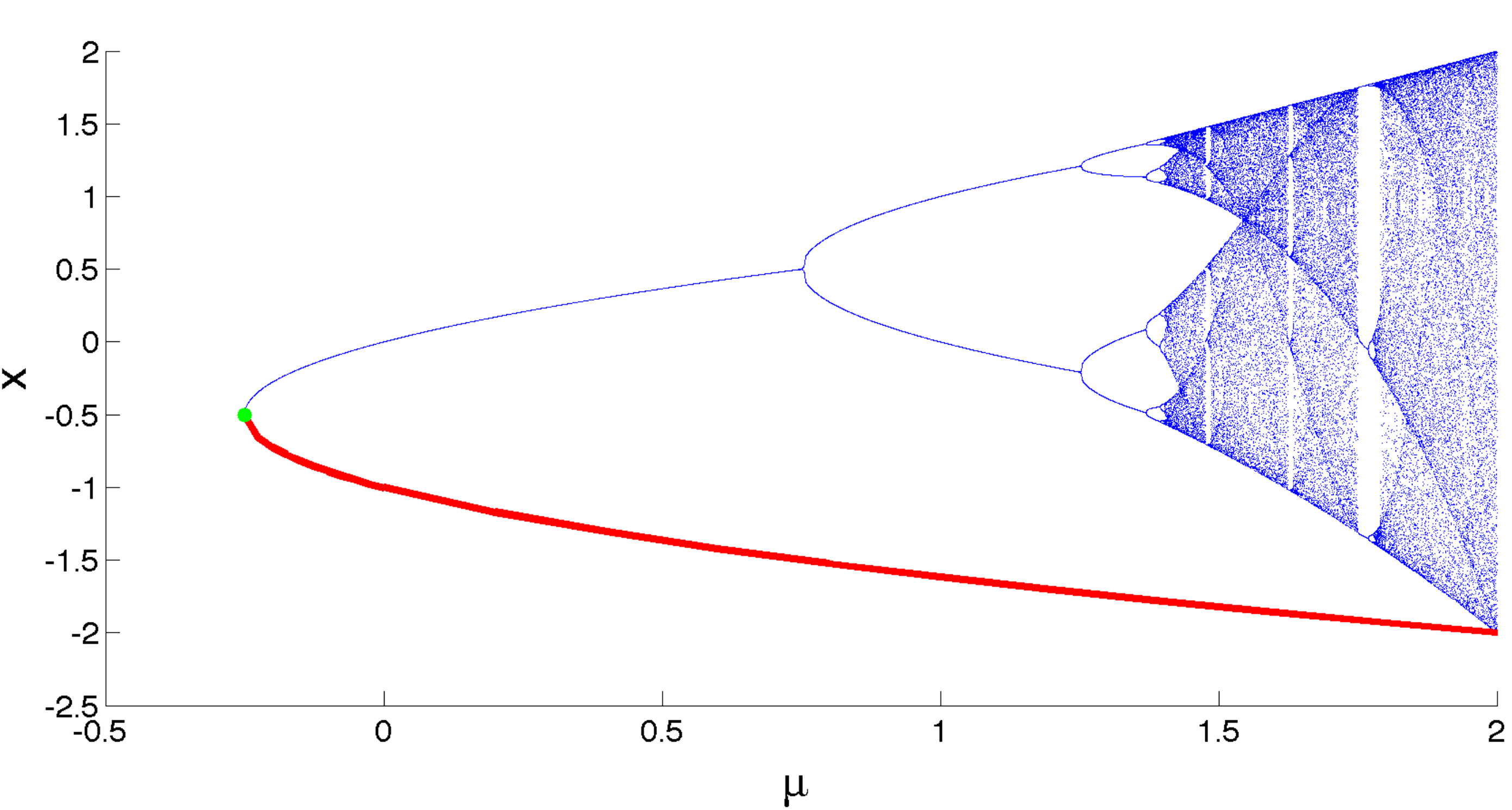}
\caption{\label{fig:mu-xx} {\bf Cascades for $F(\mu,x)=\mu - x^2$. }
  This figure shows the attracting set for $F$ for $-0.25<\mu<2$. The
  attracting set is created at a saddle-node bifurcation at
  $\mu=-0.25$ (green dot). The path of unstable fixed points (red) exists
  for all $\mu>-0.25$. The stable fixed point undergoes infinitely
  many period-doubling bifurcations, limiting to the value $\mu
  \approx 1.4$. This set of period doublings is called a
  period-doubling cascade. This map also has infinitely many 
  period-doubling cascades that begin with periodic orbits of period $> 1$. 
The red curve consists of unstable regular fixed points that exist for all $\mu > -0.25$.
}
\end{center}
\end{figure}

\begin{figure}
\includegraphics[width=\textwidth]{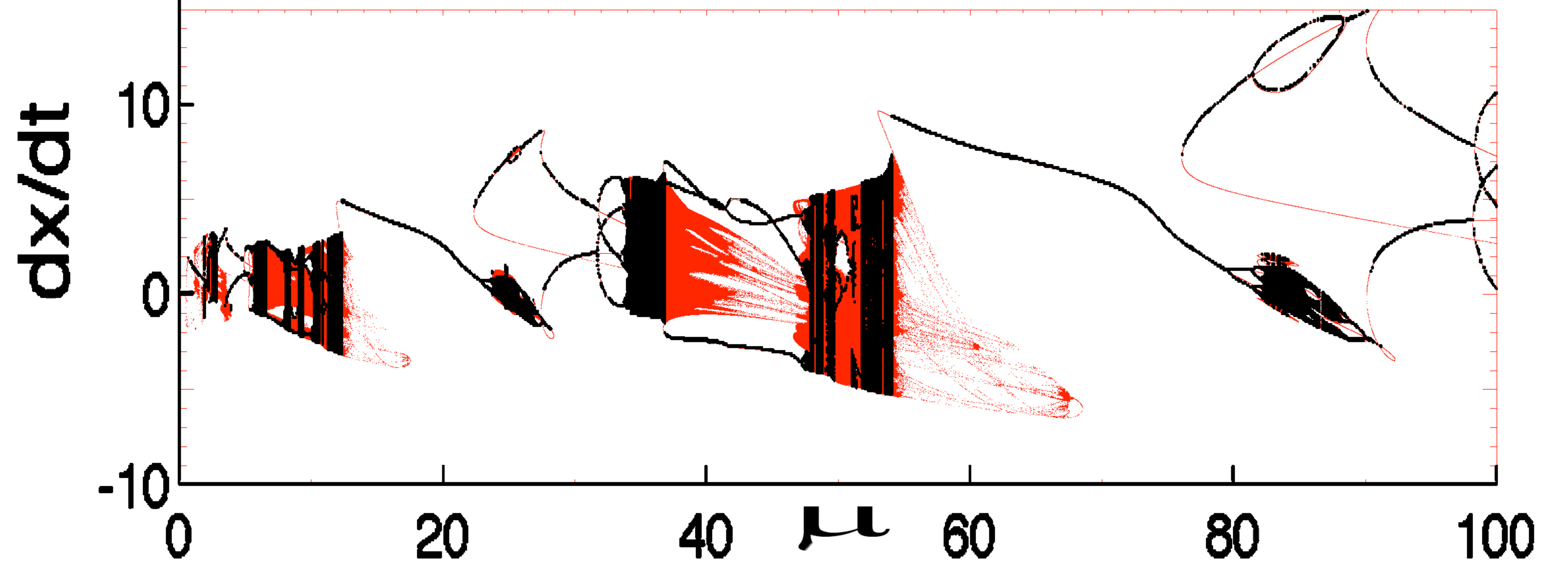}
\caption{{\bf Cascades in the double-well Duffing equation.} 
The attracting sets (in black) and periodic orbits up to period ten
(in red)
for the time-$2\pi$ map of the double-well Duffing equation: 
$x''(t)+0.3 x'(t)-x(t)+(x(t))^2+(x(t))^3=\mu \sin t$. 
Numerical studies show regions of chaos
interspersed with regions without chaos, as in the Off-On-Off Chaos
Theorem (Theorem~\ref{theorem:offonoff}). 
\label{Fig:vdp} }
\end{figure}
\begin{figure}
\includegraphics[width=\textwidth]{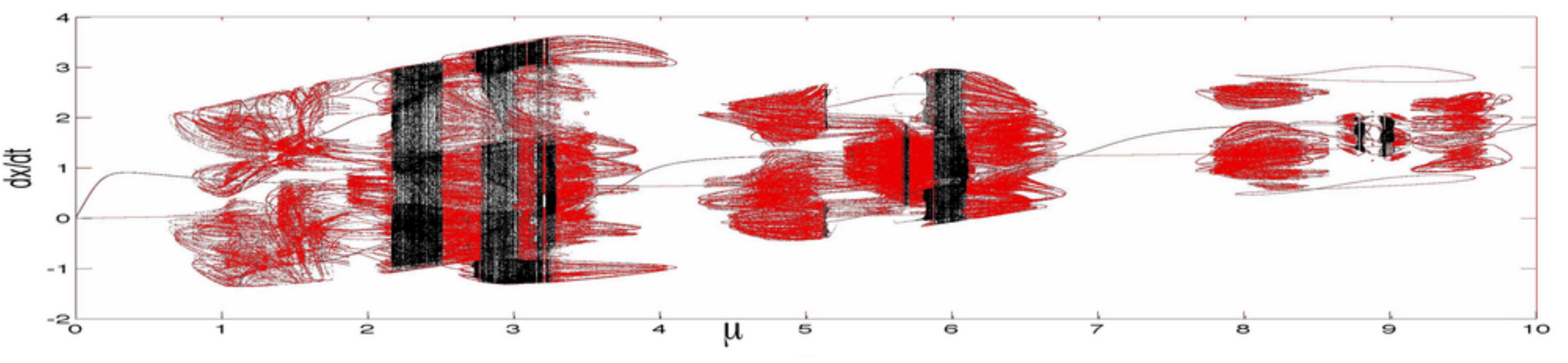}
\caption{{\bf The forced-damped pendulum.} 
For this figure, periodic points with periods up to ten were plotted 
in red for the time-$2\pi$ map of the forced-damped pendulum equation: 
$x''(t)+0.2 x'(t)+\sin(x(t))=\mu \cos(t)$, indicating the general areas with chaotic dynamics for this map.
Then the attracting sets were plotted in black, hiding some periodic points. 
Parameter ranges with and without chaos are interspersed. 
\label{Fig:pend} }
\end{figure}

In Figure ~\ref{fig:mu-xx}, as $\mu$ increases from $\mu=-0.25$ towards a
value $\mu_F \approx 1.4$, a family of periodic orbits undergoes an
infinite sequence of period doublings with the periods of these orbits
tending to $\infty$. This infinite process is called a cascade. We
will later define it more precisely. It has been repeatedly observed in
a large variety of scientific contexts that the presence of infinitely
many period-doubling cascades is a precursor to the onset of chaos. 
For example, cascades 
occur in what numerically appears as the onset to chaos for both the double-well Duffing equation, as shown Figure~\ref{Fig:vdp},
and the forced-damped pendulum, shown Figure~\ref{Fig:pend}.  Cascades were first reported by
Myrberg in 1962~\cite{myrberg:62}, and studied in more detail by
Feigenbaum~\cite{feigenbaum:79}.  This cascade is not the only
cascade. In fact, there are infinitely many distinct period-doubling
cascades. Namely, there are infinitely many windows, that is, disjoint intervals in the 
parameter that begin with a saddle-node (or source-sink) bifurcation,
and continue with the attractor undergoing an infinite sequence of period
doublings within that interval of parameters.  

\begin{figure}[t]
\begin{center}
\includegraphics[width=\textwidth]{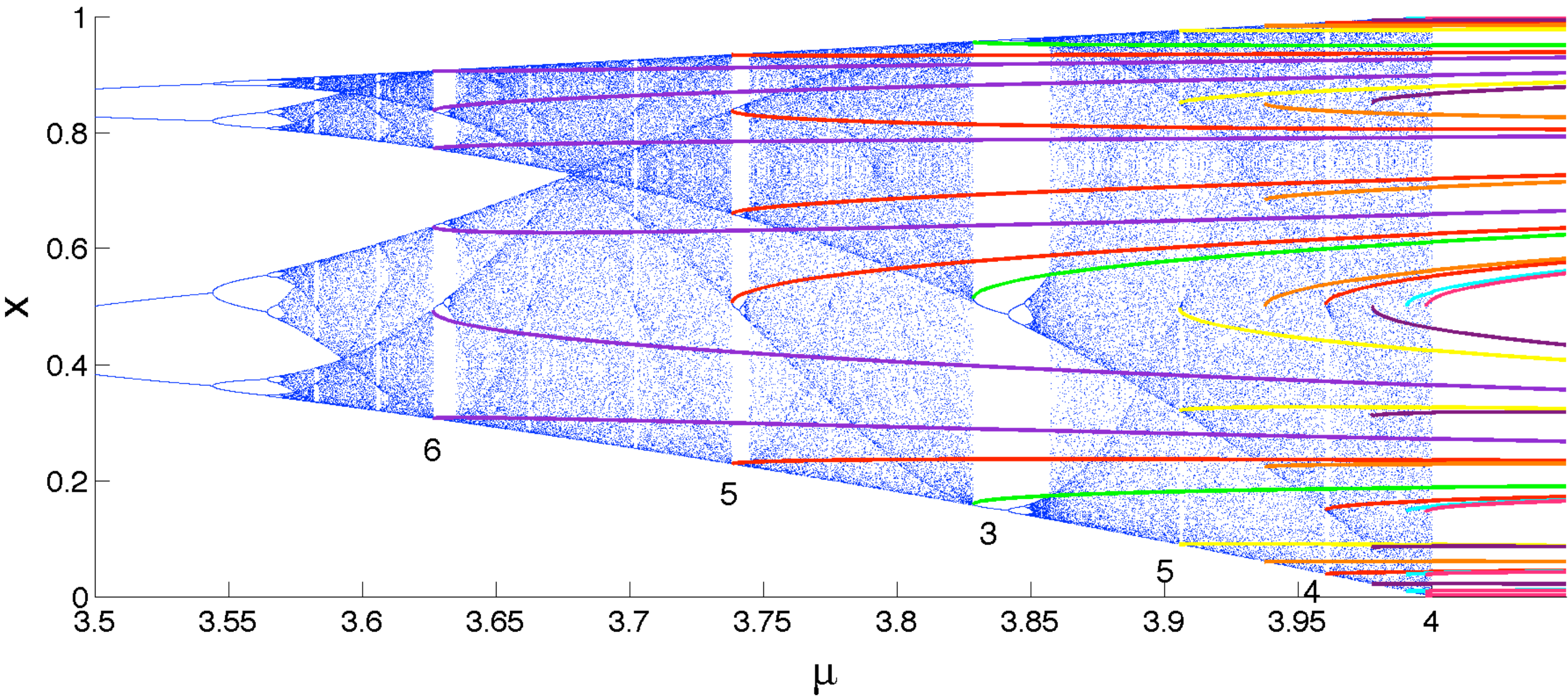}
\caption{\label{fig:quadratic} {\bf Cascades for $F(\mu,x)=\mu x
    (1-x)$.}  The logistic map has infinitely many cascades of
  attracting periodic orbits, and all cascades start at the stable
  orbit of a saddle-node bifurcation. The unstable orbits form what we
  call the {\bf stems} of the cascades (shown in color). Each stem continues to exist for all
  large $\mu$ values. By our terminology, this means that all the
  cascades shown are solitary (on any parameter interval
  $[\mu_1,\mu_2]$, for $\mu_1=3.5$ and any $\mu_2>4$) since the stem 
does not connect its cascade to a second cascade. The stems are shown here up to
  period six. Different colors are used for different periods. }
\end{center}
\end{figure}

\begin{figure}
\begin{center}
\includegraphics[width=0.9\textwidth]{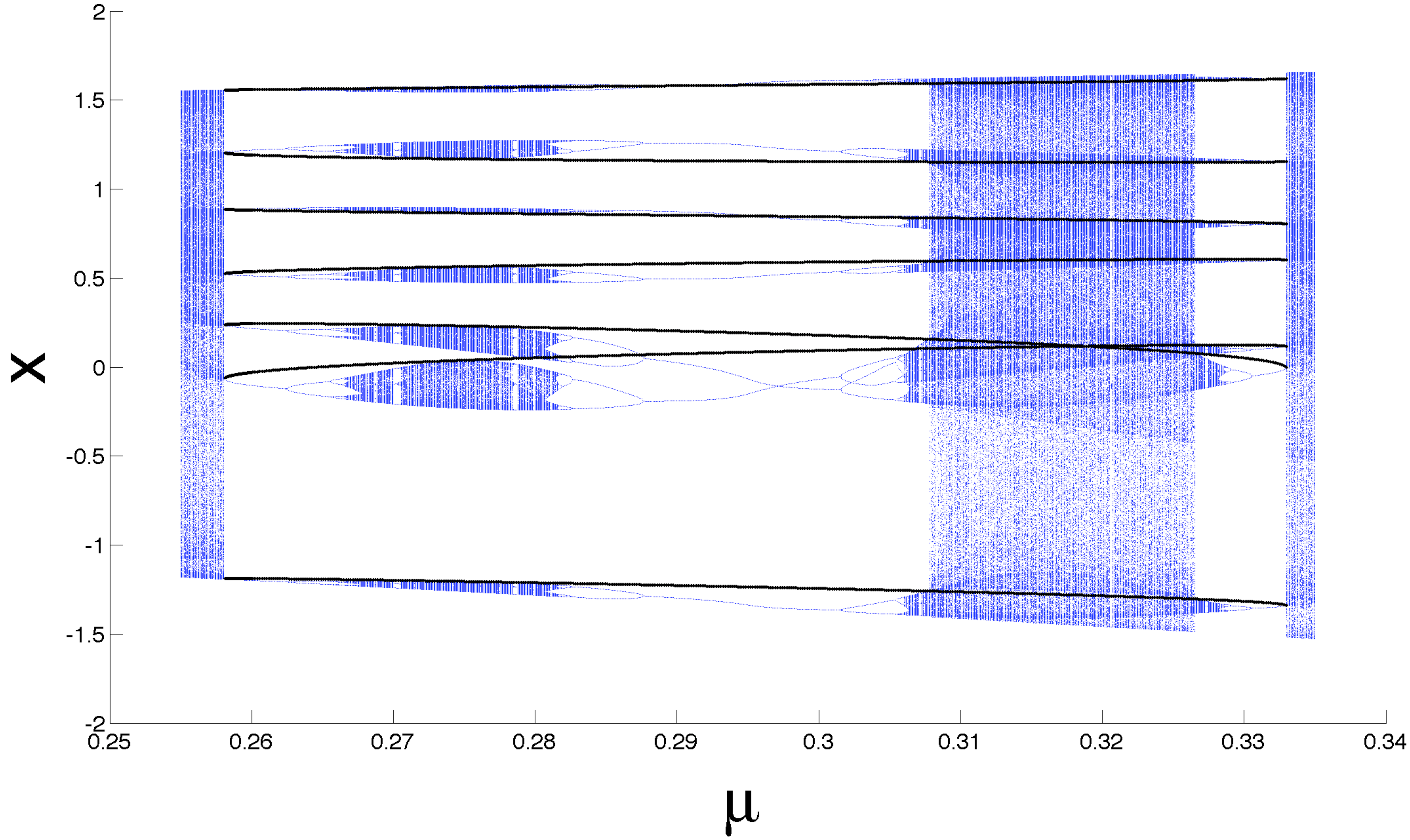}
\includegraphics[width=0.9\textwidth]{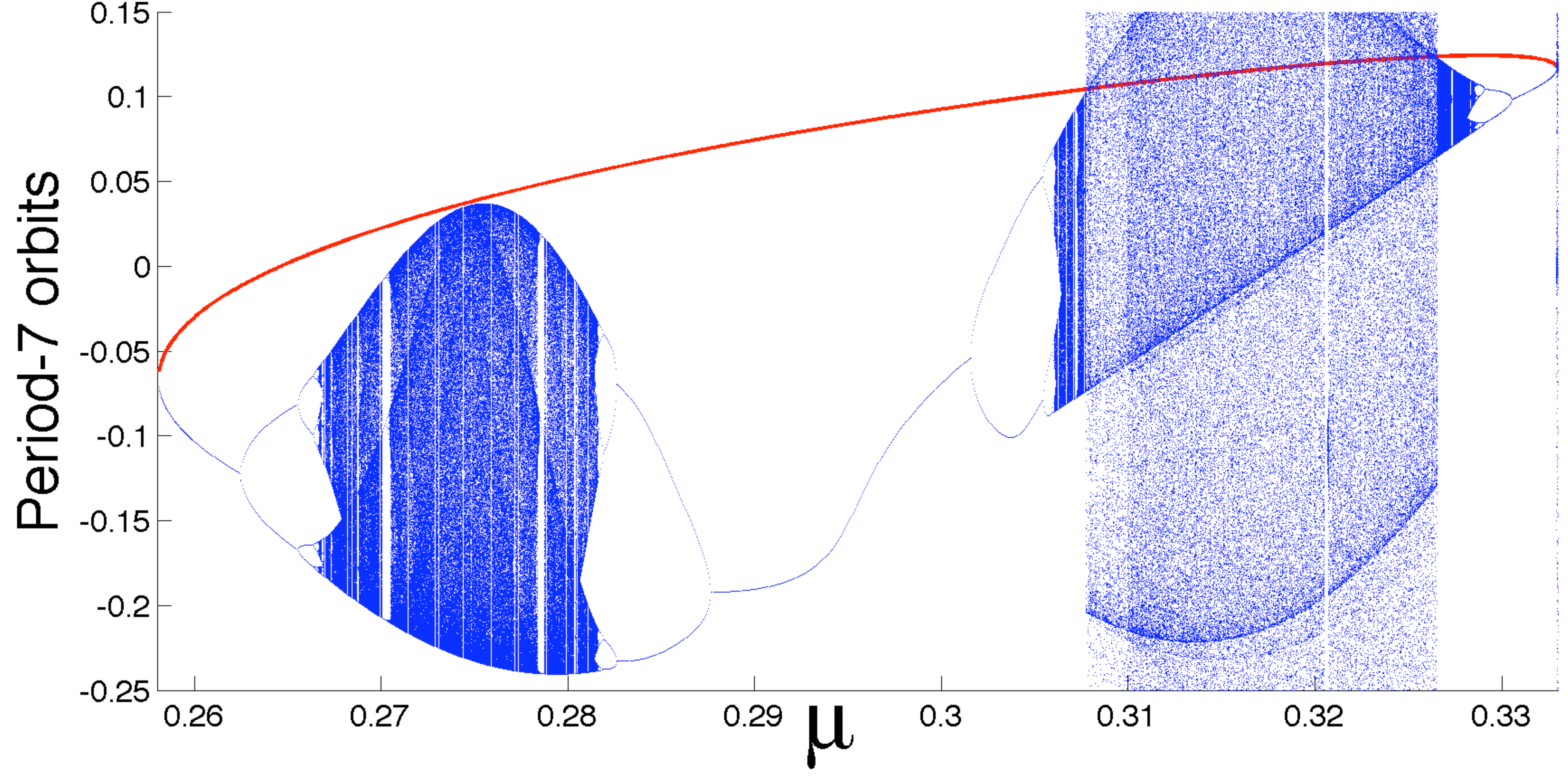}
\caption{\label{fig:Henon1} {\bf Paired cascades in the H{\'e}non
    map $(u,v)\mapsto (1.25 - u^2 + \mu v, u)$.} 
The top bifurcation diagram shows a set of four period-7 cascades.  The
  bottom bifurcation diagram shows detail of the top part. Only
  one point of each of the period-7 orbits of the H{\'e}non map are shown so
  that it is clearer how the two pairs connect to each other.  The
  leftmost and rightmost cascades form a pair that is connected by a
  path of unstable regular periodic orbits (shown in red). Likewise,
  the two middle cascades form a pair. It is connected by a
  path of attracting period-seven orbits (blue). 
Paired cascades are not robust to moderate changes in the map.}
\end{center}
\end{figure}

Quadratic maps as in the example in Figure ~\ref{fig:mu-xx}  has the quite atypical property that as the parameter
increases, there are no bifurcations that destroy periodic
orbits.  Such maps are called {\bf monotonic}. (This monotonicity was
originally proved implicitly by Douady and Hubbard in the complex analytic
setting. See the Milnor-Thurston paper \cite{milnor:thurston:88} for a proof.) 
Once one knows that a map is monotonic, it is easy to 
show that as chaos develops there
must be infinitely many cascades. See Figures ~\ref{fig:mu-xx}, and~\ref{fig:quadratic}--\ref{fig:perdoubCascade}. 

The monotonicity property is a quite severe restriction, even in
one-dimension.  No higher dimensional maps that develop chaos are
monotonic; yet numerical studies indicate that there are infinitely
many cascades whenever there is one. See for example
Figures~\ref{Fig:vdp},~\ref{Fig:pend}, and~\ref{fig:Henon1}. In this paper we summarize our progress and
give new extensions to our theory that explains why there are infinitely
many cascades in the onset to chaos. Our explanation is also valid for maps of arbitrary
dimension.

In the first result of this paper, we consider the context in which
virtually all periodic saddles have the same unstable dimension. (By
{\bf virtually} we mean all except for a finite number.) In this case,
the onset of chaotic behavior always includes infinitely many
cascades.

There is an extensive literature on Routes to Chaos; that is,
situations in which for some $\mu_1$ and $\mu_2$, the trajectory of
$x$ under $F(\mu_2, \cdot)$ is in the basin of a chaotic attractor,
whereas under $F(\mu_1, \cdot)$ it is not. Whatever might have
happened that caused this change between $\mu_1$ and $\mu_2$ is called
a route to chaos. We prefer to call these ``routes to a chaotic
attractor'' to be more specific.  There are many different routes to a
chaotic attractor. See our discussion section for a partial
enumeration. For many maps there is competition between instability
and stability. For example, the appearance of an attracting periodic
orbit as a parameter is varied may mask the chaotic dynamics, and when
the orbit becomes unstable, a chaotic attractor is likely to
appear. Hence a periodic orbit's loss of stability is one example of a
route to a chaotic attractor. This approach ignores the question we
address here: how did the chaotic dynamics arise in the first place?

{\bf Two types of cascades.}  For maps with the monotonicity property,
each cascade is {\bf solitary}, in that it is not connected to another
cascade by a path of {\bf regular} periodic orbits. These paths are the
colored {\bf stems} shown in Figure ~\ref{fig:quadratic}. See Section ~\ref{sec:definitions} for full definitions of these terms. Furthermore, the
chaos persists for all parameters larger than a certain
value. However, in many scientific contexts, it is common to see
chaotic behavior appear and then disappear as the parameter $\mu$
increases. Thus as it increases there is both a route to chaos followed by a route
away from chaos. In this situation virtually all cascades are {\bf paired}; that is, two cascades are connected by a
path of regular periodic orbits. See Figure ~\ref{fig:Henon1}. 

{\bf Solitary cascades are robust.} In our second set of results, we
show that while paired cascades can be easily created and destroyed,
solitary cascades are far more robust even in the presence of rather large perturbations of a
map. Solitary cascades usually have stems with a constant period. This
{\bf stem-period} can be thought of as the period that starts the
cascade. These ideas give quite striking results. For example for each
period $p$ the following two maps \[ Q(\mu,x)= \mu - x^2  \] and \[ \widetilde{Q}(\mu,x)= \mu - x^2 + 1000 \cos(\mu^3 + x) \] have exactly the same number of solitary cascades of stem-period $p$ -- assuming the bifurcations of the second map are {\bf generic}. Namely, we call a map generic if its periodic orbit bifurcations are all generic. See Section~\ref{sec:definitions} for the full definitions of these terms. 
We know that all the bifurcation orbits are generic for 
almost every smooth map, and that if a smooth
map is not generic, then it has infinitely many generic maps arbitrarily close to it, but unfortunately -- 
with few exceptions -- we cannot tell if a given map is generic.

The map $Q$ has no paired cascades, but the $\widetilde{Q}$ may. 
For example there is exactly one solitary cascade with
stem-period $1$ and one with stem-period $3$. These results extend to
\[ F(\mu,x) = \mu - x^2 + g(\mu,x)\]
where g is smooth (ie. infinitely
continuously differentiable) and $|g(\mu,x)|$ and $|g_x(\mu,x)|$ are
uniformly bounded -- as would be the
case if $g$ was smooth and periodic in each variable, again assuming
the map $F$ has generic bifurcations.


{\bf Outline.} The paper proceeds as follows: In Section~\ref{sec:definitions}, we
give some basic definitions, including what we mean by a cascade, the 
definition of chaos that we use here, and the class of generic maps with which we work.
Section~\ref{sec:firstresult} contains a series of results relating
chaos and cascades, with an explanation of the concrete relationship
between periodic orbits within the chaos and the resulting cascades
along the route towards this chaos.  In
Section~\ref{sec:secondresult}, we discuss the fact that all cascades
are either paired or solitary, and show that solitary cascades are 
robust under changes in the map.  

In Section~\ref{sec:paired}, we show
that if there is chaotic behavior interspersed with non-chaotic
behavior, then virtually all cascades are paired. 
It is common in scientific applications that chaos is interspersed
with orderly behavior, in what we call {\bf off-on-off chaos} (defined
formally in Section~\ref{sec:paired}). Our numerical studies
indicate that this occurs multiple times for both the forced-damped pendulum and double-well Duffing examples. 

We end with a discussion and present open
questions in Section~\ref{sec:discussion}.

\section{Definitions}\label{sec:definitions}

\begin{figure}
\begin{center}
\includegraphics[width=\textwidth]{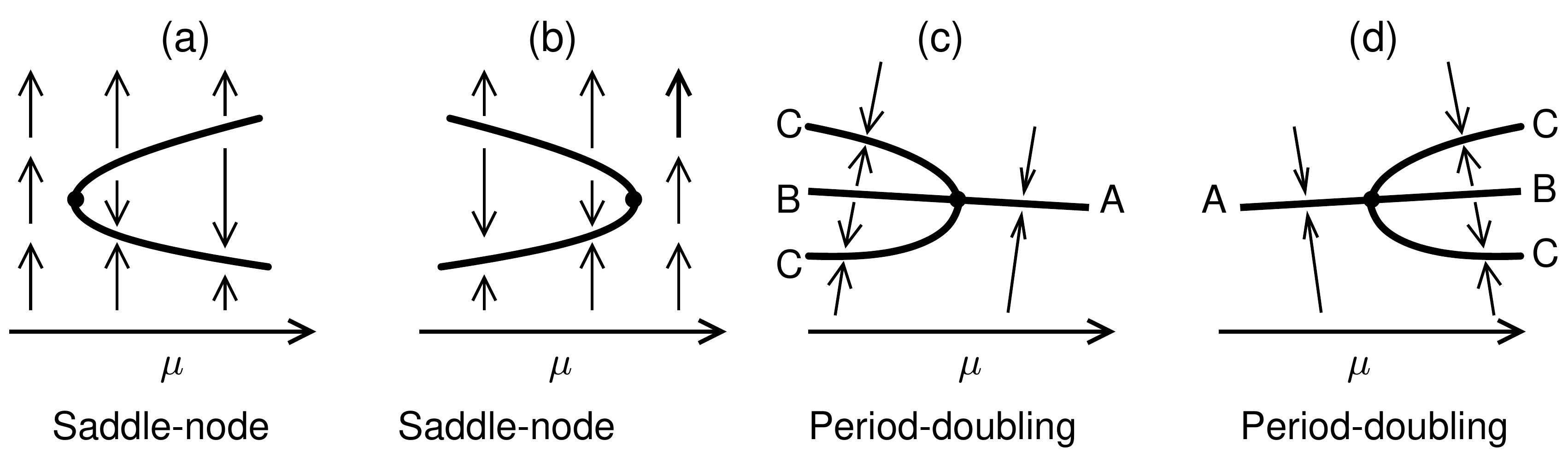}
\caption{\label{fig:direction-reversing} {\bf Center manifold of saddle-node bifurcations
    and period-doubling bifurcations.}  This figure shows typical
  saddle-node and period-doubling (halving) bifurcations, along with
  the stability. For generic maps $F:R\times R^n$, it is sufficient to examine the $R\times R$ center manifold. We plot one-dimensional $x$ vertically and $\mu$ horizontally.
We use vertical arrows to show how the stability
  varies near a bifurcation periodic point -- indicated by a large
  dot.  In each case, all the stability arrows can be reversed, thereby generating four more cases. }
\end{center}
\end{figure}

\begin{figure}[t]
\begin{center}
\includegraphics[width=\textwidth]{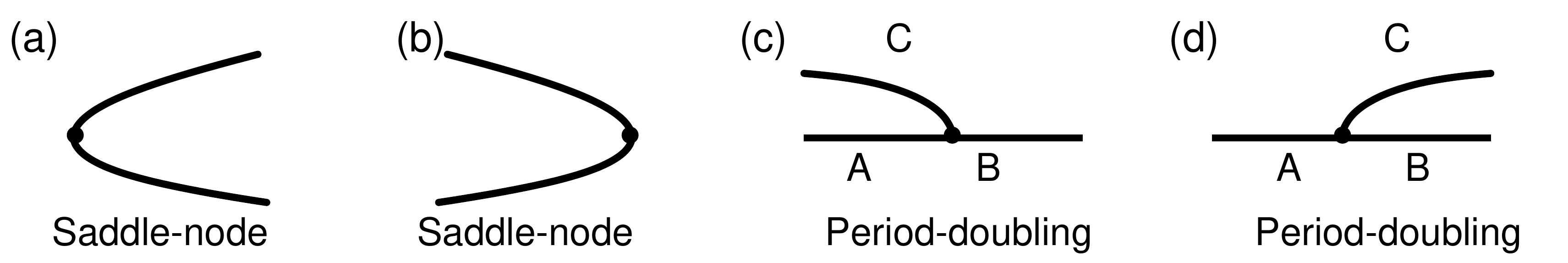}
\caption{\label{fig:snpd} {\bf The regular periodic orbits form a one-manifold
    near regular saddle-node and period-doubling bifurcation orbits.}
  In this schematic figure each point is an orbit and the horizontal
  axis is the parameter, usually $\mu$ in this paper.  (a \& b): Near a
  standard saddle-node bifurcation of a periodic orbit, the local
  invariant set consists of a curve of periodic orbits. They are
  either all flip orbits or all regular periodic orbits.  Therefore
  RPO is locally a curve. (c \& d): Near a standard period-doubling (or
  period-halving) bifurcation of a periodic orbit, the local invariant
  set consists of two curves of periodic orbits, one of period $p$
  shown as segment AB, and one of period $2p$ shown as segment
  C. Segment C always consists of regular periodic orbits, whereas
  exactly one of A and B consists of flip orbits and the other regular
  periodic orbits. Thus RPO is locally a curve consisting of C and
  either A or B, depending on which is regular.  For the quadratic map
  $\mu - x^2$, only (a) and (d) occur. That is, periodic orbits are
  created but never destroyed as $\mu$ increases. When $x$ is
  two-dimensional, such simplicity virtually never
  occurs~\cite{yorke:antimonotonicity-Annals92}.  }
\end{center}
\end{figure}

\begin{figure}
\begin{center}
\includegraphics[width=.5\textwidth]{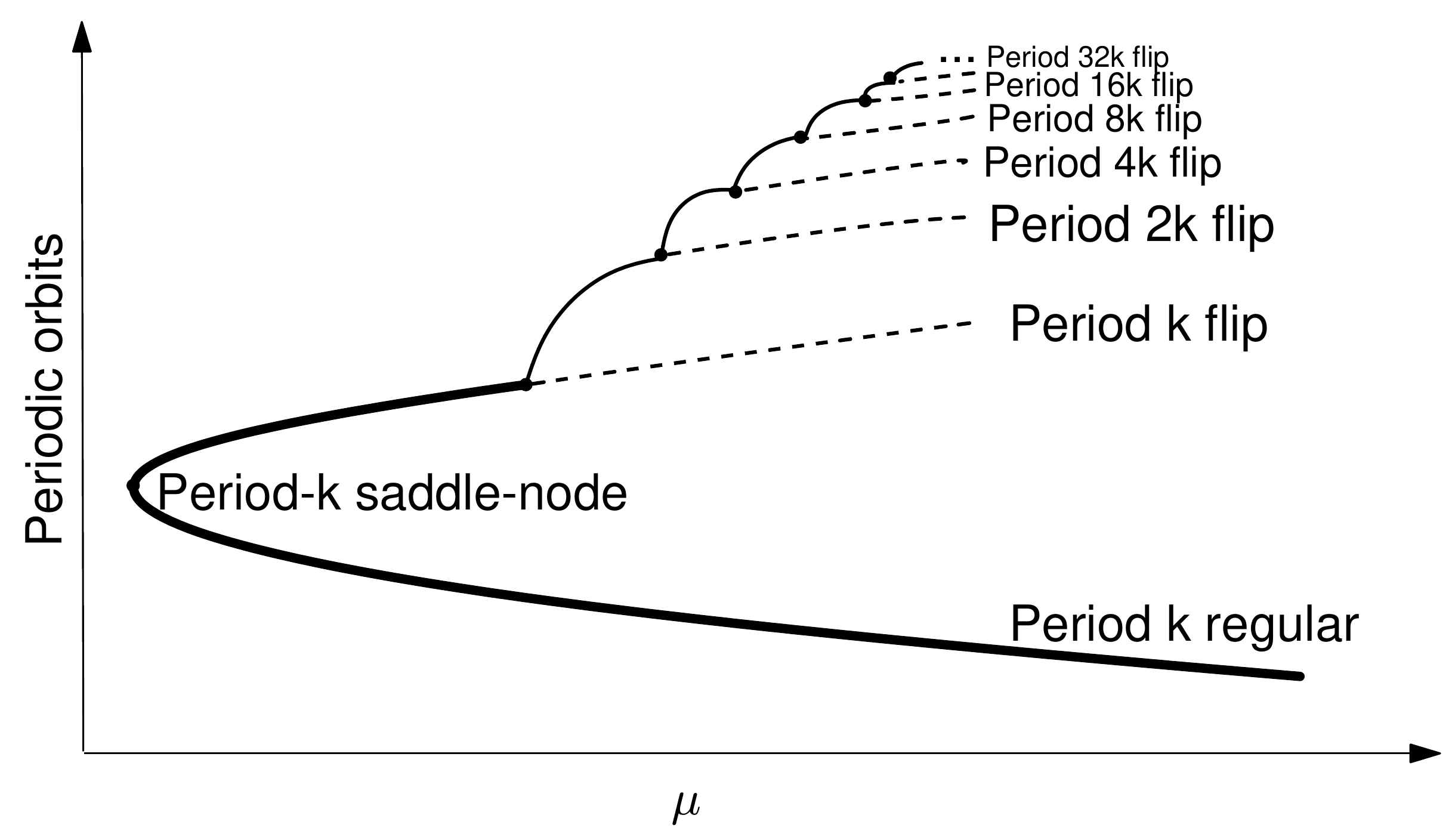}
\caption{\label{fig:perdoubCascade} {\bf A depiction of a 
    monotonic cascade.}  A cascade is a path of regular periodic orbits that has
  infinitely many period-doubling bifurcations with the periods going
  to infinity at the end of the path. Solid lines denote regular
  orbits, and dashed lines denote flip orbits.  This figure uses only
  the orbit-creation bifurcations, (a) and (d) in Figure~\ref{fig:snpd}. If only bifurcation types (a) and (d) are present and $x$ is scalar, then each saddle-node spawns an unstable branch and an attractor as $\mu$ increases. For $\mu
  - x^2$, there are no attractors for $\mu > \mu_2$ for $\mu_2$
  sufficiently large. Hence this branch of attractors cannot continue forever and
  must bifurcate. Only (d) is available, spawning a flip orbit branch
  and a doubled-period branch of attractors. Again this new branch of
  attractors cannot continue forever as $\mu$ increases. In this way,
  an infinite set of period-doublings results before $\mu$ reaches
  $\mu_2$.  Much more complicated patterns are possible when all four types of bifurcations are allowed, including
  period-halving as well as period-doubling, and the path may contain
  many saddle-node bifurcations. }
\end{center}
\end{figure}

We investigate smooth maps $F(\mu, x)$ where $\mu$ is in an interval
$J$, and $x$ is in a smooth manifold $\mathfrak M$ of any finite
dimension. For example, for the forced damped pendulum,
\[ \frac{d^2 \theta}{dt^2} + A \frac{d\theta}{dt} + \sin
\theta = \mu \cos t, \]
we take $F(\mu, x)$ to be the time-$T$ map, where $T= 2\pi$ is the
period of the forcing, and $x=(\theta, d\theta /dt)$. Then the first
coordinate of $x$ is on a circle, and the second is a real number. Hence
$\mathfrak M$ is a cylinder. \\

We say a point $(\mu,x_0)$ is a {\bf period-$p$ point} if $F^p(\mu,x_0)=
x_0$ and $p$ is the smallest positive integer for which that is
true. Its {\bf orbit},
sometimes written $[ ( \mu,x_0 ) ]$, 
is the set
\[ \{ (\mu, x_0), (\mu, x_1), \cdots, (\mu, x_{p-1}) \}, \mbox{ where } x_j =
F^j(\mu,x_0).\]

By the {\bf eigenvalues} of a period-$p$ point $(\mu,x_0)$, we mean the
eigenvalues of the Jacobian matrix $D_xF^p(\mu,x_0)$.

An orbit is called {\bf hyperbolic} if none of its eigenvalues has
absolute value 1. All other orbits are {\bf bifurcation
  orbits}. Figure~\ref{fig:direction-reversing} depicts two standard examples of
bifurcation orbits and the resulting stability of nearby periodic
points.

We call a periodic orbit a {\bf flip} orbit if the orbit has an odd
number of eigenvalues less than -1, and -1 is not an eigenvalue. (In
one dimension, this condition is: derivative with respect to $x$ is $<-1$. In dimension two,
flip orbits are those with exactly one eigenvalue $< -1$.) All other
periodic orbits are called {\bf regular}.  For example, the periodic
orbits of constant period switch between flip and regular orbits at a
period-doubling bifurcation orbit since an eigenvalue crosses
$-1$. See Figure~\ref{fig:snpd}. We write {\bf RPO} for the set of
regular periodic orbits.

For some $a,b \in R$ and $\psi \in [a, b)$, let $ Y(\psi) =
(\mu(\psi), x(\psi))$ be a {\bf path of regular periodic points}
depending continuously on $\psi$. Assume $\psi$ does not retrace any orbits. That is,
each $Y(\psi)$ is a periodic point on a regular periodic orbit, and
distinct values of $\psi$ correspond to periodic points on distinct
orbits.  We call a regular path $Y(\psi)$ a {\bf cascade} if the path
contains infinitely many period-doubling bifurcations, and for some
period $p$, the periods of the points in the path are precisely $p,
2p, 4p, 8p, \dots $. As one traverses the cascade, the periods need
not increase monotonically, but as $\psi \to b$, the period of
$Y(\psi)$ goes to $\infty$.  The orbits of a cascade with monotonic period increase
are depicted schematically in Figure~\ref{fig:perdoubCascade}.

Write $fixed(\mu,p)$ for the set of fixed points of $F^p(\mu, \cdot)$
and $|fixed(\mu,p)|$ for the {\it number} of those fixed
points.  We say that there is {\bf exponential periodic orbit growth}
at $\mu$ if there is a number $G > 1$ for which the number of periodic
orbits of period p satisfies $|fixed(\mu,p)| \ge G^p$ for infinitely
many $p$. For example, this inequality might hold for all even $p$, but for
odd $p$ there might be no periodic orbits. This is equivalent to the
statement that for some $h(= \log G)>0 $, we  have 
\begin{equation} \label{eqn:limsup} 
\mbox{limsup}_{p \to \infty} \frac{\log |fixed(\mu,p)|}{p} \ge h. \end{equation}

\bigskip
\noindent
{\bf Periodic orbit chaos.}
Chaotic behavior is a real world phenomenon, and trying to give it a
single definition is like trying to define what a horse
is. Definitions are imperfect. A child's definition of a horse might
be clear but would be unsatisfactory for a geneticist (whose
definition might be in terms of DNA) and neither would satisfy a
breeder of horses who might give a recursive definition, ``an
offspring of two horses.'' Definitions of real-life phenomena describe
aspects of that phenomena. They might agree in the great majority of
cases on which animals are horses, though there may be  rare atypical exceptions like clones where they might
disagree. Just as it is impossible currently to connect the shape and
sound of a horse with its DNA sequence, it is similarly impossible
currently to identify in full generality positive Lyapunov exponents
with exponential growth of periodic orbits.

Similarly ``chaos'' and ``chaotic'' should have definitions
appropriate to the needs of the user. On the other hand, an
experimenter might insist that to be chaotic, there must be a chaotic
attractor, until he/she starts looking for chaos on basin boundaries
and finds transient chaos. That approach leaves no terms for the chaos
that occurs outside an attractor, as on fractal basin boundaries. We
make no such restriction. Our results involve periodic orbits, and we make our 
definition accordingly.

We say a map $F(\mu, \cdot)$ has {\bf periodic orbit (PO) chaos} at a
parameter $\mu$ if there is exponential periodic orbit growth.  This
occurs whenever there is a horseshoe for some iterate of the map. It
is sufficiently general to include having one or multiple co-existing
chaotic attractors, as well as the case of transient chaos. As hinted at by Equation~\ref{eqn:limsup}, in many
cases PO chaos is equivalent to positive topological entropy. We discuss this
relationship further in Section~\ref{sec:discussion}.

The {\bf unstable
  dimension} Dim$_u(\mu, x_0)$
of a periodic point $(\mu,x_0)$ or periodic orbit is defined to be the number of its eigenvalues
$\lambda$ having $|\lambda| > 1$, counting multiplicities.

We say there is {\bf virtually uniform (PO) chaos} at $\mu$ if there is PO chaos, and
all but a finite number of periodic orbits have the same unstable
dimension, denoted Dim$_u(\mu)$.

For the pendulum map discussed above, whenever there is PO chaos at
some parameter value $\mu$, we expect the periodic orbits to be
primarily saddles, and if it likewise had virtually uniform PO chaos,
then we would expect $Dim_u(\mu) = 1$.  Assuming
there are infinitely many periodic orbits, roughly half would be
regular saddles, with the rest being flip saddles.
Furthermore, all attracting periodic points are regular. 

Our first goal is to describe the {\bf route to (PO) chaos}. That is, if at
$\mu_1$ there is no chaos, while at $\mu_2$ there is virtually uniform chaos, we
explain what must happen in this interval in order for chaos to arise. 

We believe that generally there is one typical route to chaotic
dynamics. Namely, there must be infinitely many period-doubling cascades 
when $\mu$ is between $\mu_1$ and  $\mu_2$. (Each of these cascades in turn has infinitely
many period-doubling bifurcations.)

\begin{figure}[htb]
\begin{center}
\includegraphics[height=.26\textwidth]{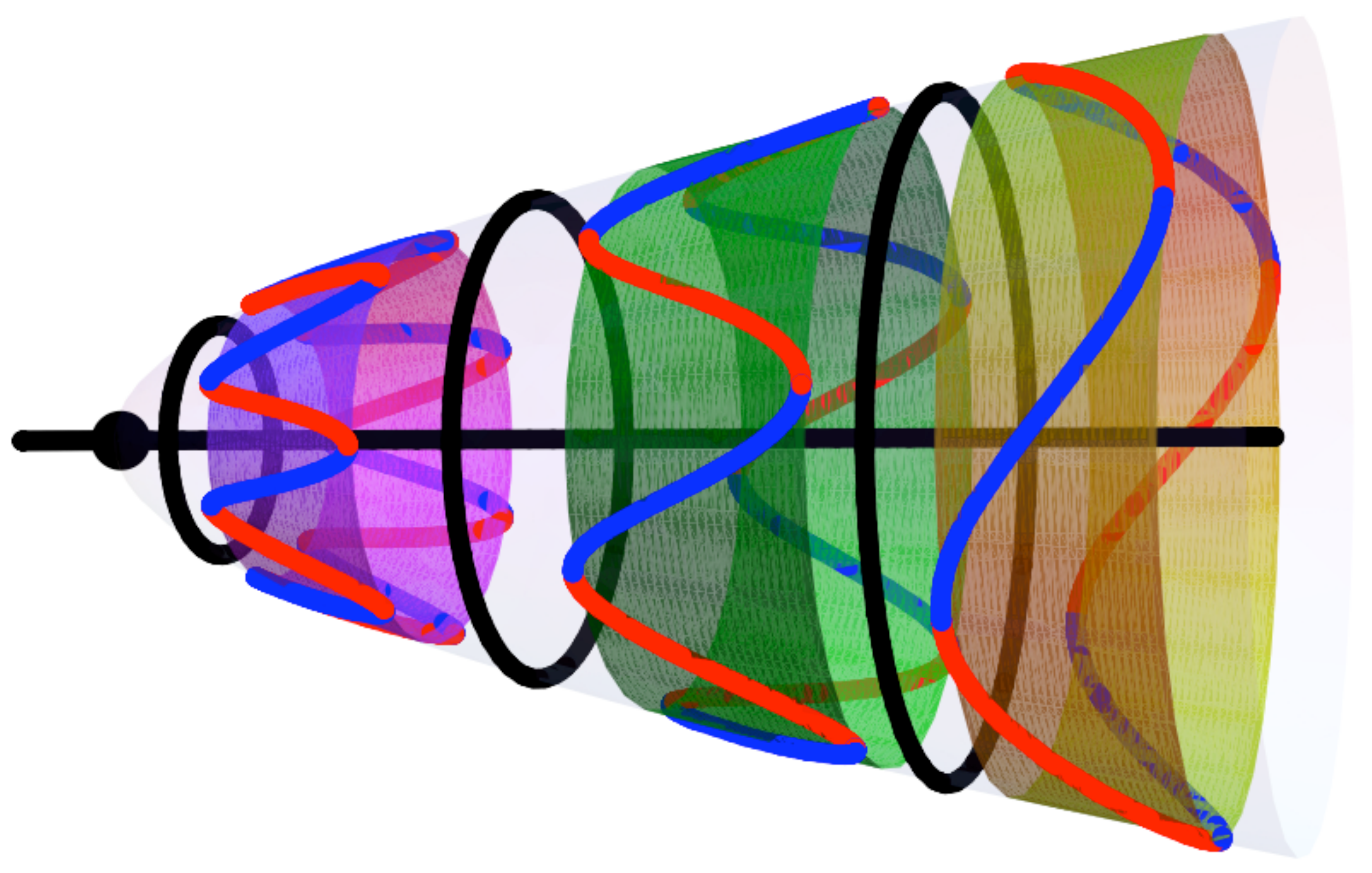}
\includegraphics[height=.26\textwidth]{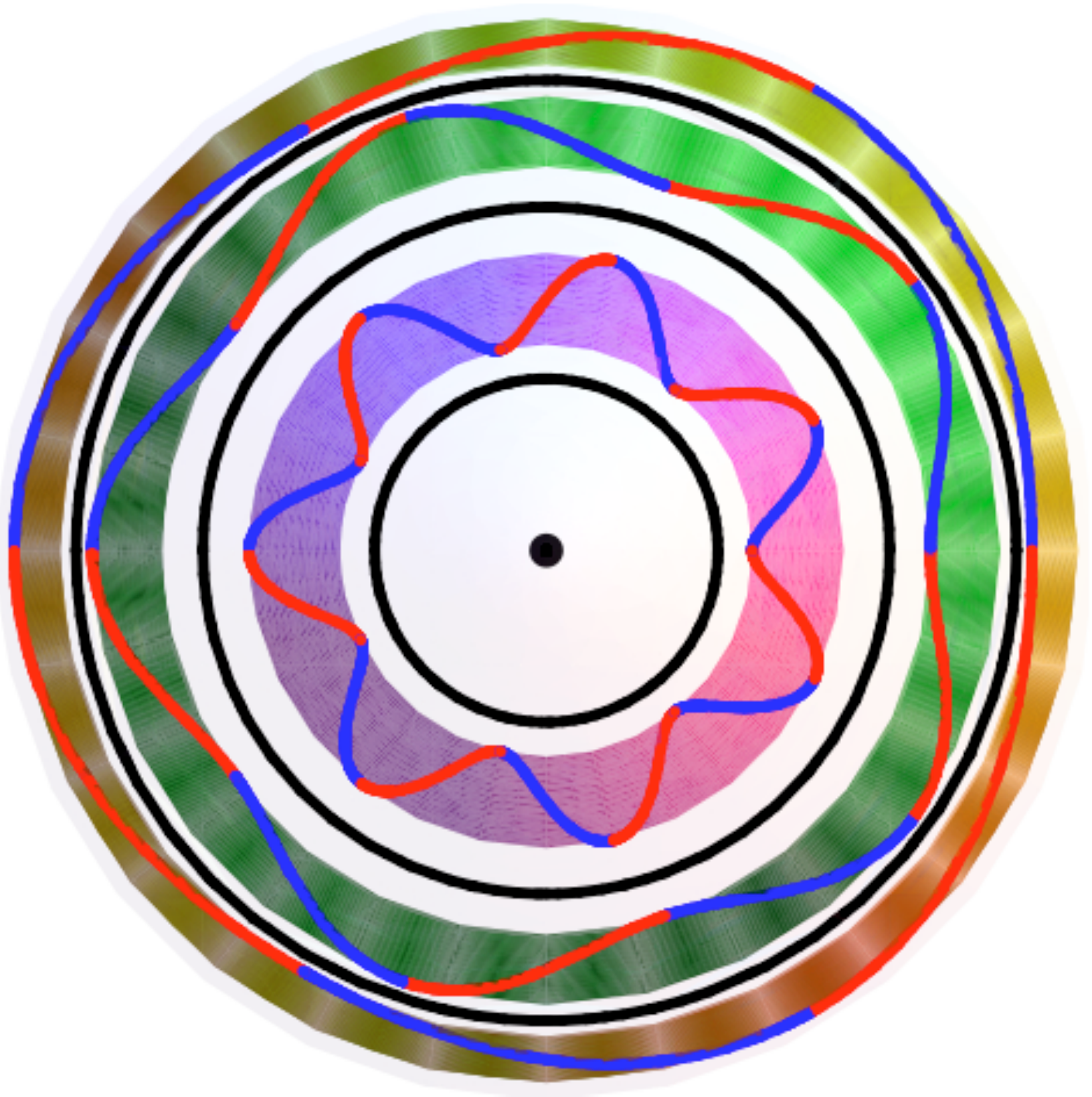}
\caption{\label{fig:hopf} {\bf A regular periodic Hopf bifurcation
    point is locally contained in a unique curve of regular periodic orbits.} 
To understand generic Hopf bifurcations, it is sufficient to determine the dynamics on its (invariant) center manifold, which is locally like $R^2 \times R$. It contains all periodic orbits that are near the bifurcation orbit.
Near a
  generic Hopf bifurcation with no eigenvalues which are roots of
  unity, the local invariant set consists of a curve of 
  periodic points (in solid black) and a surrounding paraboloid (at left).
   Although there are infinitely many periodic orbits on the paraboloid in each neighborhood
  of the Hopf bifurcation point, the middle curve of periodic points is
  disconnected from all periodic points on the paraboloid. Specifically, the paraboloid
  contains infinitely many annuli converging to the bifurcation point,
  each with periodic orbits formed and destroyed at saddle-node
  bifurcations. Typical annuli are depicted here by colored
  regions. The periodic orbits are depicted in blue (saddles) and red
  (nodes).  Between every two annuli, there is an invariant circle (in
  black) with irrational rotation number, as made clear by the
  projection (at right) of the paraboloid to the center manifold plane of the phase space.  
   }
\end{center}
\end{figure}

\bigskip
\noindent 
{\bf Generic maps.}
Our results are given for generic maps of a parameter. Specifically,
we say that the map $F$ is {\bf generic} if all of the bifurcation
orbits are {\bf generic}, meaning that each bifurcation orbit is one
of the following three types:
\begin{enumerate}
\item A standard saddle-node bifurcation. (Where ``standard''
  means the form of the bifurcation stated in a standard textbook,
  such as Robinson~\cite{robinson:95}.) In particular the orbit has
  only one eigenvalue $\lambda$ for which $|\lambda|=1$, namely
  $\lambda=1$.

\item A standard period-doubling bifurcation. In particular the orbit
  has only one eigenvalue $\lambda$ for which $|\lambda|=1$, namely
  $\lambda=-1$.

\item A standard Hopf bifurcation. In particular the orbit has only
  one complex pair of eigenvalues $\lambda$ for which
  $|\lambda|=1$. We require that these eigenvalues are not roots of
  unity; that is, there is no integer $k > 0$ for which $\lambda^k=1$.
\end{enumerate}

These three bifurcations are depicted in Figures~\ref{fig:snpd}
and~\ref{fig:hopf}. Generic $F$ have at most a countable number of
bifurcation orbits, so almost every $\mu$ has no bifurcation
orbits. See~\cite{sander:yorke:p09} for the details showing that these
maps are indeed generic in the class of smooth one-parameter
families. For systems with symmetry such as the forced-damped
pendulum, a fourth type of bifurcation occurs, such as a pitchfork or
symmetry-breaking bifurcation. This adds complications, though in fact
with extra work our results remain true.

Our motivation for considering generic maps is given in
Proposition~\ref{prop:components} in the next section, which states
that each regular periodic orbit for a generic map is locally
contained in a unique path of periodic orbits. The connection to
cascades can be summarized as follows: starting at each regular
periodic orbit $Q$ for $\mu \in [\mu_1,\mu_2]$, there is a local path
of regular periodic orbits through $Q$. Enlarge this path as far 
as possible. Either the path reaches $\mu_1$ or $\mu_2$, or there is a
cascade. This idea is explained in more detail in the next section. 

\section{Onset of chaos implies cascades}\label{sec:firstresult}

Our first result is Theorem~\ref{theorem:version1}, which demonstrates
that the route to virtually uniform PO chaos contains infinitely many
period-doubling cascades. Theorem~\ref{theorem:version2} is a
restatement of these results in a way that makes the relationship between chaos and cascades much more
transparent.

Write $J$ for the closed parameter interval $[\mu_1, \mu_2]$. Our main
hypotheses will be used for a variety of results so we state them
here.

\bigskip
\noindent
{\bf List of Assumptions.}
\begin{enumerate}
\item[($A_0$)] Assume $F$ is a generic smooth map; that is, $F$ is
  infinitely differentiable in $\mu$ and $x$, and all of its
  bifurcation orbits are generic.
\item[($A_1$)] Assume there is a
  bounded set $M$ that contains all periodic points $(\mu, x)$ for
  $\mu \in J$.
\item[($A_2$)] Assume all periodic orbits at $\mu_1$ and $\mu_2$ are hyperbolic.
\item[($A_3$)] Assume that the number $\Lambda_1$ of  periodic orbits at $\mu_1$ is finite. 
\item[($A_4$)] Assume at $\mu_2$ there is virtually uniform PO chaos. Write
  $\Lambda_2$ for the number of  periodic orbits at $\mu_2$ having
  unstable dimension not equal to Dim$_u(\mu_2)$.
\end{enumerate}

\begin{theorem} \label{theorem:version1}
Assume ($A_0~-~A_4$).
Then there are infinitely many
distinct period-doubling cascades between $\mu_1$ and $\mu_2$.
\end{theorem}

\begin{example} 
Based on numerical
studies, a number of maps appear to satisfy the conditions of the
above theorem. Note that numerical verification
involves significantly more work than just plotting the attracting sets
for each parameter, since we are concerned about both the stable and
the unstable behavior to determine whether there is chaos. Examples
include the time-$2\pi$ maps for the double-well Duffing (Fig.~\ref{Fig:vdp}), 
the triple-well Duffing, the forced-damped pendulum (Fig.~\ref{Fig:pend}), 
the Ikeda map (introduced to describe the
field of a laser cavity), and the pulsed damped rotor map.
\end{example}

This formulation gives no idea how the behavior at $\mu_2$ is
connected to the cascades that must exist. Thus before giving a proof,
we reformulate the conclusions of the theorem in a more transparent
way.  Specifically, we reformulate so that it is clear how infinitely many
cascades in the strip $S = [\mu_1,\mu_2] \times \mathfrak M$ are
connected to regular periodic orbits at $\mu_2$ by continuous paths in
RPO.

\bigskip
\noindent
{\bf Paths of orbits.}
We now give a formal definition for paths of regular periodic orbits and show that they connect regular periodic orbits at
$\mu_2$ with cascades between $\mu_1$ and $\mu_2$. We start with the
following theorem, which guarantees that a path through each regular
periodic orbit is unique.

Assume hypotheses $A_0-A_4$, and consider a local path through a
regular periodic orbit as guaranteed by the above proposition.
Specifically, let $Y_0 = (\mu_2, x_0)$ be a regular periodic point.
Since all such points at $\mu_2$ in the above theorem are assumed to
be hyperbolic, the orbit can be followed without a change of direction
for nearby $\mu< \mu_2$. We can define $Y(\psi) = (\mu(\psi),
x(\psi))$ continuously for $\psi$ in an interval $[\psi _0, \psi _1]$
as follows.  Let $Y(\psi_0) = (\mu_2, x_0)$. Then $Y(\psi)$ follows
the branch of regular periodic points for decreasing $\mu$ until
reaching $Y(\psi_1)$, which is either a bifurcation orbit or
$\mu = \mu_1$. If it is a bifurcation point, which would be a generic
bifurcation point, then from the above proposition there is a unique
branch of regular hyperbolic periodic points leading away from
$Y(\psi_1)$, and $Y(\psi)$ follows that branch. If $Y(\psi_1)$ is a
period-doubling bifurcation point, there are two branches of regular
periodic points, but both are on the same orbits. Either branch can be
followed since both represent the same orbits and both are regular.

Each branch of hyperbolic periodic orbits can be parametrized by $\mu$
and is easy to find numerically by solving a differential equation.
Write $A(\mu)$ for the square matrix $D_xF^p(\mu,x(\mu))$ and $b(\mu)$
for the vector $D_{\mu}F^p(\mu,x(\mu))$. Differentiating

\[ F^p(\mu,x(\mu)) = x(\mu)\]
 with respect to $\mu$ and manipulating yields
\[b(\mu) + A(\mu) \frac{dx}{d\mu} = \frac{dx}{d\mu}, \mbox{ or } \] 
\[\frac{dx}{d\mu} = -(A(\mu) - Id)^{-1} \; b(\mu),\]

Where $Id$ denotes the identity matrix. The matrix inverse above
exists since $(\mu,x(\mu))$, being hyperbolic, cannot have $+1$ as an eigenvalue.

This construction suggests a
definition. We say $Y(\psi)$ is a {\bf path} in $S$ for $\psi$ in some
interval $K$ if it is continuous and 
\begin{itemize}

\item[(i)] $Y(\psi)$ is a regular periodic point in $S$ for each $\psi \in K$;

\item[(ii)] Y does not retrace orbits; that is, $Y$ is never on the same
orbit for different $\psi$.
\end{itemize}

\begin{proposition}[Local paths of regular periodic points]\label{prop:components}
  For a smooth generic map $F: R \times {\mathfrak M} \to {\mathfrak
    M}$, each regular periodic point is locally contained in a path of
  regular periodic points. The corresponding path of periodic orbits is unique.
\end{proposition}

This proposition appears in reference~\cite{sander:yorke:p09}.  Here we
give an idea of the proof.  The implicit function theorem shows
that there is a local curve of periodic points through a regular
hyperbolic periodic point. Therefore, the proof consists of showing
that there is also a local curve of orbits through each periodic bifurcation orbit. This must be shown for
each of the three types of generic bifurcations. The curve is not 
necessarily
monotonic: It may reverse direction with respect to $\mu$, as depicted
in Figure~\ref{fig:direction-reversing}. Namely, a path $ Y(\psi) =
(\mu(\psi), x(\psi))$ {\bf reverses direction} at $\psi_0$ if
$\mu(\psi)$ changes from increasing to decreasing or vice versa at
that point.  For n-dimensional $x$, the study of generic saddle-node
and period-doubling bifurcations can be reduced to one-dimensional $x$
due to the center manifold theorem.  The bifurcation orbit has $n-1$
eigenvalues outside the unit circle, and these do not affect the
dynamics on the two-dimensional invariant $(\mu,x)$-plane. Hence here
we can consider $x$ as being one-dimensional.

In the case of a saddle-node bifurcation, one branch of orbits is
stable and one is unstable in $R^1$.  Since the other $n-1$
eigenvalues do not cross the unit circle near the point, it follows
that the unstable dimension of the orbits in $n-$dimensions is even
for one branch and odd for the other, i.e., the unstable dimension
$k(\psi)$ of $Y(\psi)$ changes parity as $Y(\psi)$ passes through the
bifurcation point $Y(\psi_0)$.

For a period-doubling bifurcation, consider the notation depicted in
Figure~\ref{fig:direction-reversing}. Namely, let (A) denote the period-$n$ orbits
with no coexisting period-$2n$ orbits, let (B) denote the branch of
period-$n$ orbits coexisting at the same parameter values with the
branch (C) of period-$2n$ orbits.  If branch (A) is regular and (B) is
flip, then the path $Y$ includes branches (A) and (C), which are on
different sides of the bifurcation point, and $\mu(\psi)$ does not
change direction and $k(\psi)$ does not change parity. If however (B)
is regular and (A) is flip, then $\mu(\psi)$ does change direction and
$k(\psi)$ does change parity.

For Hopf bifurcations (see Figure~\ref{fig:hopf}), the path proceeds
monotonically past the bifurcation point and the unstable dimension
changes by $\pm 2$, so the parity does not change.  Hence in all
cases, the path changes direction if and only if the parity changes. The cases of saddle-node
and period-doubling bifurcations are straightforward, as depicted in
Figure~\ref{fig:snpd}.  To show that RPO is locally a curve at a Hopf
bifurcation is trickier than the other two cases, since there are in
fact infinitely-many periodic orbits near a Hopf bifurcation point,
but they are not connected to the Hopf bifurcation point by any path
of periodic points, as shown in Figure~\ref{fig:hopf}. The proof
follows from the steps listed.

\bigskip

\noindent {\bf Cascades.}
We call a regular path $Y(\psi)$ for $\psi\in [a,b)$ a {\bf cascade}
if the path contains infinitely many period-doubling bifurcations, and
for some period p, the periods of the points in the path are precisely
$p, 2p, 4p, 8p, \cdots $. As one traverses the cascade, the periods
need not increase monotonically, but as $\psi \to b$, the period of
$Y(\psi)$ goes to $\infty$.

We will say a path $Y(\psi)$ is {\bf maximal} if the following additional condition holds:
\begin{itemize}
\item[(iii)] $Y$ cannot be extended further to a larger interval, and it cannot be
redefined to include points of more regular orbits.
\end{itemize}

\begin{figure}[t]
\begin{center}
\includegraphics[width=.45\textwidth]{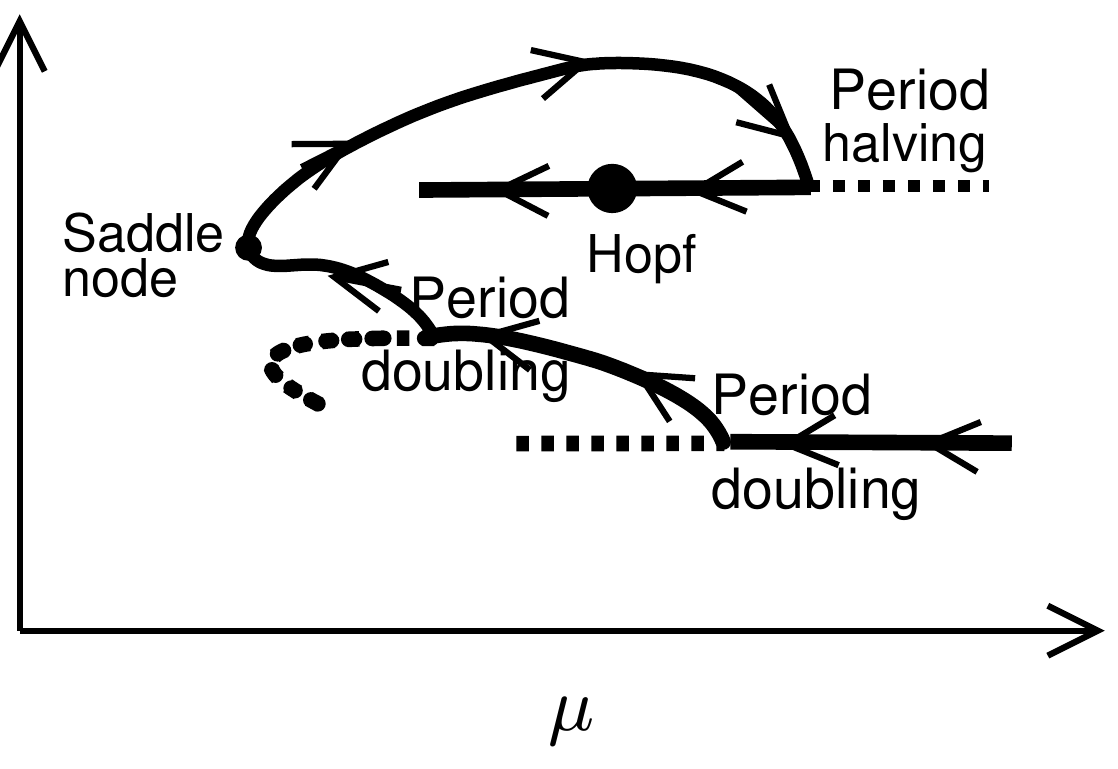}
\caption{\label{fig:maximal-path} {\bf Paths of regular periodic orbits.}  We
  assume here that all bifurcation orbits are generic. It follows that
  a regular periodic point is in a path of periodic points.  Since
  each periodic point is in a periodic orbit,  we can think of the
  path as being a path of regular periodic orbits. This is a unique
  path of regular orbits (solid).  Here we show orbits, not
  points. Flip orbits are depicted by dashed lines, showing that the
  path is no longer unique if we include all periodic orbits. Let
  $Y(\psi)$ be a path of regular (non-flip) periodic points
  parametrized by $\psi$, such that $Y(\psi)$ does not pass through the same orbit
  twice. Write $[Y(\psi)]$ for the orbit that the point $Y(\psi)$ is
  in. It is a useful fact that whenever the $\mu$ coordinate
  $\mu(\psi)$ of $Y(\psi)$ changes direction, the unstable dimension
  of $Y(\psi)$ changes by one. Hence it changes from odd to even or
  vice versa. The arrows along the path show which way the path is
  traveling as $\psi$ increases.}
\end{center}
\end{figure}

\begin{figure}
\begin{center}
\includegraphics[width=.45\textwidth]{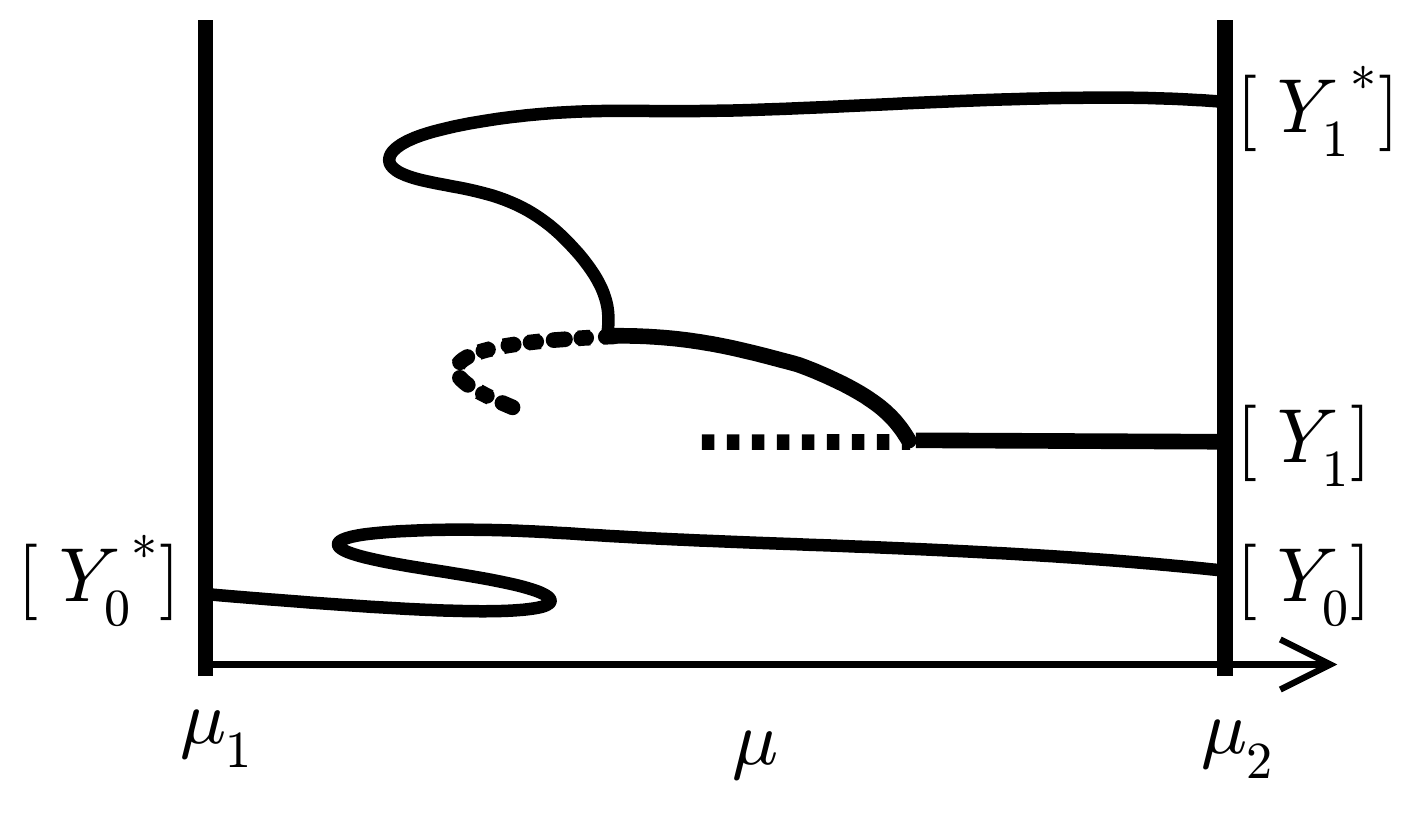}
\caption{\label{fig:PathsFromChaos1} {\bf Paths of regular periodic orbits
    starting from $\mu_2$.}  Assume the bifurcations
  are generic. We use the notation of the above figure.  Let $Y_0 =
  (\mu_2, x_0)$ be a regular hyperbolic periodic point. If its
  maximal path of orbits extends to a hyperbolic orbit $Y_0^*$ at
  $\mu_1$, then the path has changed directions an even number of
  times and the unstable dimensions of $Y_0$ and $Y_0^*$ have the same
  parity; that is, both are even or both are odd. If the path starting
  at a regular hyperbolic orbit $Y_1 = (\mu_2, x_1)$ returns to a
  point $Y_1^*$ at $\mu_2$, the path changes directions an odd number
  of times, and so the unstable dimensions of $Y_1$ and $Y_1^*$ have
  opposite parity and in particular the unstable dimensions of the two
  are different. }
\end{center}
\end{figure}

\begin{figure}[t]
\begin{center}
\includegraphics[width=.45\textwidth]{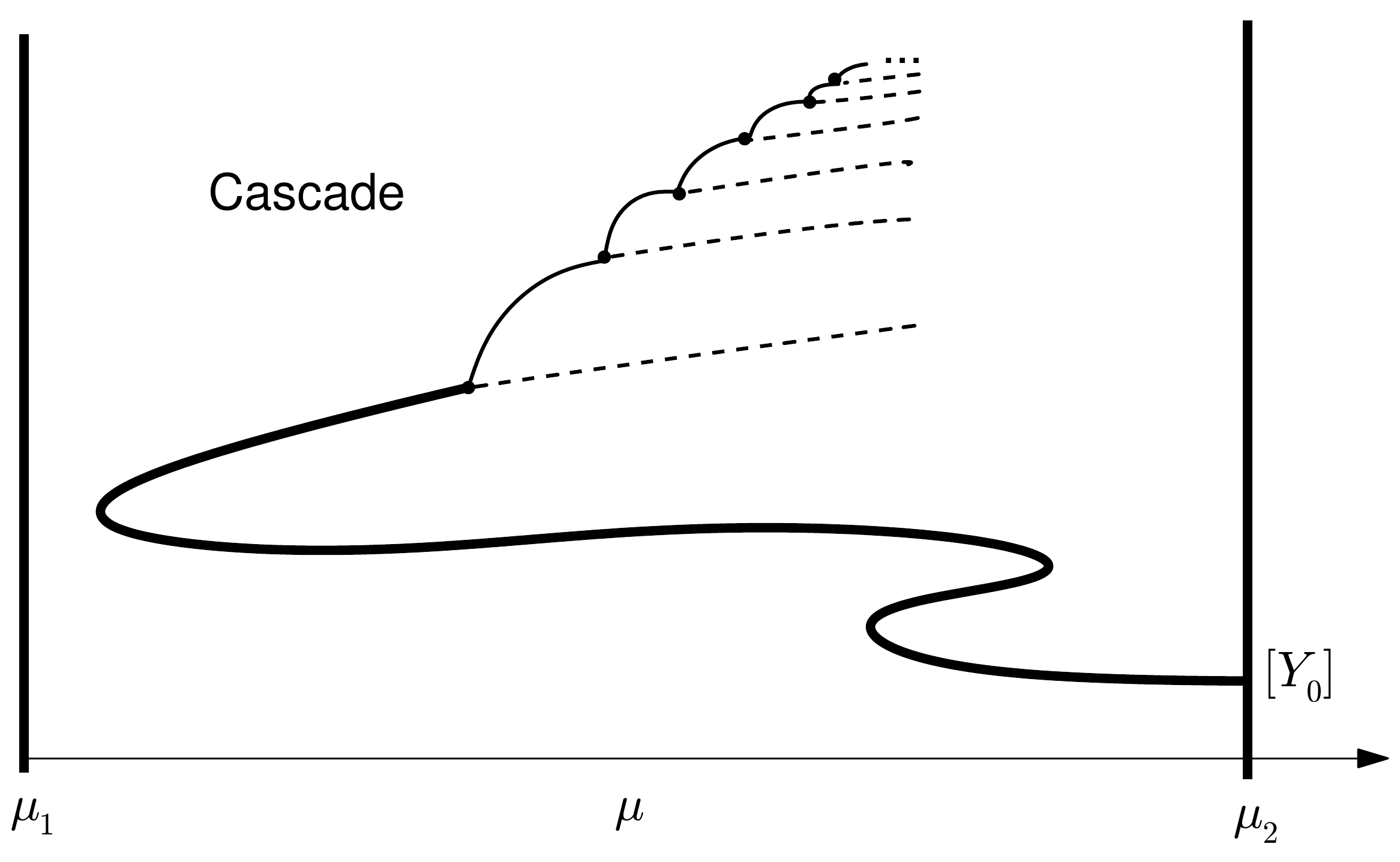}
\caption{\label{fig:PathsFromChaos2} {\bf An interior path of regular orbits
    starting from $\mu_2$ yields cascades.}  Continuing the assumptions and notations
  of the above figures, assume the hypotheses of
  Theorem~\ref{theorem:version1}. Only a finite number of paths that
  start at $\mu_2$ can either reach $\mu_1$ or return to
  $\mu_2$. Hence there are an infinite number of regular hyperbolic
  orbits at $\mu_2$ that yield paths that neither reach $\mu_1$ nor
  return to $\mu_2$. Such {\bf interior paths} must contain an infinite
  number of bifurcations. Since generically there are only a finite
  number of bifurcation orbits of each period between $\mu_1$ and
  $\mu_2$, the period of the orbits must tend to infinity along the
  path and it must contain a cascade. Hence there are infinitely many
  cascades between $\mu_1$ and $\mu_2$.}
\end{center}
\end{figure}

Figure~\ref{fig:maximal-path} shows an example of a middle portion of a maximal
path. Figure~\ref{fig:PathsFromChaos1} and~\ref{fig:PathsFromChaos2} show
different possibilities for how the maximal path ends. 

Let Orbits(Y) be the set of periodic points on orbits traversed by
$Y$. That is, if $(\mu, x) = Y(\psi)$ for some $\psi$, then $(\mu, x)
\in \mbox{Orbits}(Y)$ and so are the other points on the orbit, $(\mu,F^n(\mu, x))$ for all n.

Two integers are said to have the same {\bf parity} if both are odd or
both are even. Otherwise they have opposite
parity. Figures~\ref{fig:direction-reversing} and 
\ref{fig:maximal-path} demonstrate why this is a
critical idea. Namely, the unstable dimension of a path $Y(\psi)$ {\bf changes
  parity} as $\psi$ increases precisely when the path changes
directions. Thus the parity of the unstable dimension corresponds to
the orientation of the path.

\begin{theorem} \label{theorem:version2}
Assume ($A_0~-~A_4$). Then the following are true.
\begin{itemize}
\item[(B1)] there are infinitely many regular periodic points at $\mu_2$.

\item[(B2)]  
For each maximal path $Y(\psi) = (\mu(\psi),x(\psi))$ in $S$ starting
from a regular periodic point $Y_0 = (\mu_2,x_0)$, the set of orbits traversed, denoted by
Orbits$(Y)$, depends only on the initial orbit containing
$Y_0$.  That is, different initial points on the same orbit
yield paths that traverse the same set of orbits, so we can write
Orbits$(Y_0)$ for Orbits$(Y)$.

\item[(B3)] Let $Y_0 = (\mu_2,x_0)$ and $Y_1=(\mu_2,x_1)$ be regular periodic points on
different orbits. Then Orbits$(Y_0)$ and Orbits$(Y_1)$ are
disjoint.

\item[(B4)] Let $K$ denote the unstable dimension of a regular periodic point
$(\mu_2,x_0)$. For a maximal path $Y(\psi) = (\mu(\psi),x(\psi))$ in
$S$ starting from $(\mu_2,x_0)$, let $k(\psi)$ denote the unstable
dimension of $Y(\psi)$.  At each direction-reversing bifurcation,
$k(\psi)$ changes parity; that is it changes from odd to even or
vice versa. Initially $Y(a) = (\mu_2,x_0)$ so initially $\mu(\psi)$ is
decreasing and $k(a) = K$, so initially $k(a)+K$ is even. Hence in
general $\mu(\psi)$ is decreasing if $K+k(\psi)$ is even and 
increasing if it is odd.

\item[(B5)] Let $Y$ be a maximal path on $[a,b]$ in $S$ and $\mu(a) =
\mu_2$. If $\mu(b) = \mu_2$, then $\mu(\psi)$ is increasing at that point
so $k(a) + k(b)$ is even. Hence $k(a) + k(b)$ is odd, so $k(a) \ne
k(b)$.

\item[(B6)] There are infinitely many distinct period-doubling cascades
between $\mu_1$ and $\mu_2$.

\item[(B7)] There are at most $\Lambda = \Lambda_1 + \Lambda_2$ regular
periodic orbits at $\mu_2$ with unstable dimension Dim$_u(\mu_2)$ that
are not connected to cascades.
\end{itemize}
\end{theorem}

Theorem~\ref{theorem:version2} is a version of
Theorem~\ref{theorem:version1} that allows us to show that most
regular orbits at $\mu_2$ are connected to cascade by a path of
regular orbits. The conclusions of this theorem were shown to hold
in~\cite{sander:yorke:p09} under assumptions $A_0-A_4$ along with the
additional hypothesis that at $\mu_2$, there are infinitely many
periodic regular orbits. However, we are now able to show that this
last hypothesis is unnecessary because it is automatically true. In
fact, under assumptions $A_0-A_4$, approximately half of the periodic
orbits are regular. This proof is quite technical in that it involves
topological fixed point index theory. To make this treatment more
readable, we give the full details in~\cite{sander:yorke:p10b}.  

\begin{example} Assume phase space is two-dimensional. If at $\mu_2$ there is
a transverse homoclinic point then there will be infinitely many
saddles and exponential orbit growth. Condition $A_2$ is often
satisfied if the attractor(s) are periodic orbits and there is a Smale
horseshoe in the dynamics.
\end{example}

\begin{example} Our numerical studies of the forced-damped pendulum
  indicate that it satisfies the hypotheses of this theorem on
  $[\mu_1,\mu_2] \approx [1.8,20], [73,20], [73,175],$ and
  $[350,175]$. Thus there are infinitely-many cascades on each of
  these intervals.  
\end{example}

\bigskip
\noindent
{\bf Weakening the uniformity hypothesis.}
We end this section with the conjecture that $A_4$ can be weakened
without effecting the conclusions of the theorem.  Specifically, let
$\Phi(\mu, K ,p)$ be the fraction of all period-p orbits at $\mu$ that
have unstable dimension $K$.  We will say that {\bf most periodic
  orbits have unstable dimension $K$} if $\Phi(\mu, K,p) \to 1$ as $p
\to \infty$.  We believe that for most generic smooth maps and most
$\mu$, there is some dimension $u_0$ for which this holds.

\begin{conjecture}
  Assume ($A_0~-~A_3$) (not including ($A_4$)).  Assume that most
  periodic orbits at $\mu_2$ have unstable dimension $K$.  Then there
  are infinitely many regular periodic orbits with unstable dimension
  $K$, and infinitely many of these are connected to distinct
  period-doubling cascades between $\mu_1$ and $\mu_2$.
\end{conjecture}

\section{Conservation of solitary cascades}\label{sec:secondresult}

Let $J$ be an interval, and let $Y(\psi)$ for $\psi \in J$ be a
maximal path of regular periodic orbits containing a cascade.  For any
point $s_0$ in the interior of $J$, infinitely many period
doublings of the cascade occur either for $\psi < s_0$ or $\psi >
s_0$.  In the following definition, we distinguish two types of
cascades based on what happens to the other end of $Y(\psi)$. 

\begin{definition}
  Let $F$ satisfy $A_0-A_2$. A cascade is {\bf solitary} in
  $[\mu_1,\mu_2]$ if the maximal path of regular periodic orbits
  containing it contains no other cascades. In this case, we call the
  non-cascade end of the maximal path the {\bf stem} of the cascade. A
  cascade is {\bf paired} if its maximal path contains two
  cascades. A cascade is {\bf bounded} in $[\mu_1,\mu_2]$ if its
  maximal path  is contained in the interior of the interval and never hits the
  boundary $\mu=\mu_1$ or $\mu=\mu_2$. We call a path which stays in
  the interior of the interval an {\bf interior path}.
\end{definition}

On a sufficiently large parameter interval, all cascades for quadratic
maps are solitary, as shown in
Figure~\ref{fig:quadratic}. Figure~\ref{fig:Henon1} shows an example
of two sets of paired cascades. The following theorem shows that the
classification of solitary versus paired is equivalent to classifying
cascades in an interval as being purely interior versus hitting the
interval boundary.

\begin{theorem}[Solitary and paired cascades]\label{theorem:solitarypaired}
  Let $F$ satisfy $A_0-A_2$ on $[\mu_1,\mu_2]$.  A cascade is paired
  if and only if the maximal path containing the cascade is an interior
  path. In particular, {\bf bounded cascades are always paired}.

  The maximal path of a solitary cascade contains a periodic
  point on the interval boundary; i.e. there is a point $(\mu,x)$ in
  the maximal path of the cascade such that $\mu=\mu_1$ or
  $\mu=\mu_2$. Namely, {\bf solitary cascades are never bounded}. 
\end{theorem}

\begin{proof}
  We give a sketch of the proof. Let $J$ be an interval, and let $Y(\psi)$
  for $\psi \in J$ be a maximal path of regular periodic points in
  $[\mu_1,\mu_2]$ containing a cascade as $\psi$ increases. Fix $s_0
  \in J$, and consider the one-sided maximal paths $Z_\pm(\psi)
  \subset Y(\psi)$ where $Z_\pm(\psi)=Y(\psi)$ respectively for $\psi>s_0,
  \psi<s_0$, $\psi \in J$. We have assumed that $Z_+(\psi)$ contains
  the cascade. Therefore $Z_{+}(\psi)$ is an interior path, since
  otherwise there would not be infinitely many period-doubling
  bifurcations.

  Key Fact: By the methods described in Section~\ref{sec:firstresult}, a
  one-sided maximal path of regular periodic points is an
  interior path if and only if it contains a cascade.

  Assume the cascade in $Y(\psi)$ is paired. Then $Z_-(\psi) \subset
  Y(\psi)$ contains a cascade, so by the key fact above, $Z_-(\psi)$ is an
  interior path, implying $Y(\psi)$ is as well.

  Conversely, if $Y(\psi)$ is an interior path, then $Z_-(\psi)$ is as
  well, and by the key fact $Z_-(\psi)$ contains a cascade, meaning
  that the cascades in $Z_\pm(\psi)$ are paired.
\end{proof}

\bigskip
\noindent
{\bf Conservation of solitary cascades.}
If $F$ satisfies $A_0-A_4$, and $\Lambda_1=\Lambda_2=0$, then
Theorem~\ref{theorem:version1} implies that there is {\bf
  conservation} of solitary cascades.  Specifically, let $\tilde{F}$
be any generic map such that $F$ and $\tilde{F}$ agree at $\mu_1$ and
$\mu_2$, though they may have completely different behavior inside the
interval. Then $F$ and $\tilde{F}$ have exactly the same solitary
cascade structure. Since the number of cascades is infinite, this does
not appear to be a meaningful statement. However, we can classify
solitary cascades by looking at the period of their stems, and it is
in this sense that their cascades are the
same. In Figure~\ref{fig:quadratic} we label five solitary cascades, one of period three, one of period four, two of period five, and one of period six.

\begin{figure}
\begin{center}
\includegraphics[width=.24\textwidth]{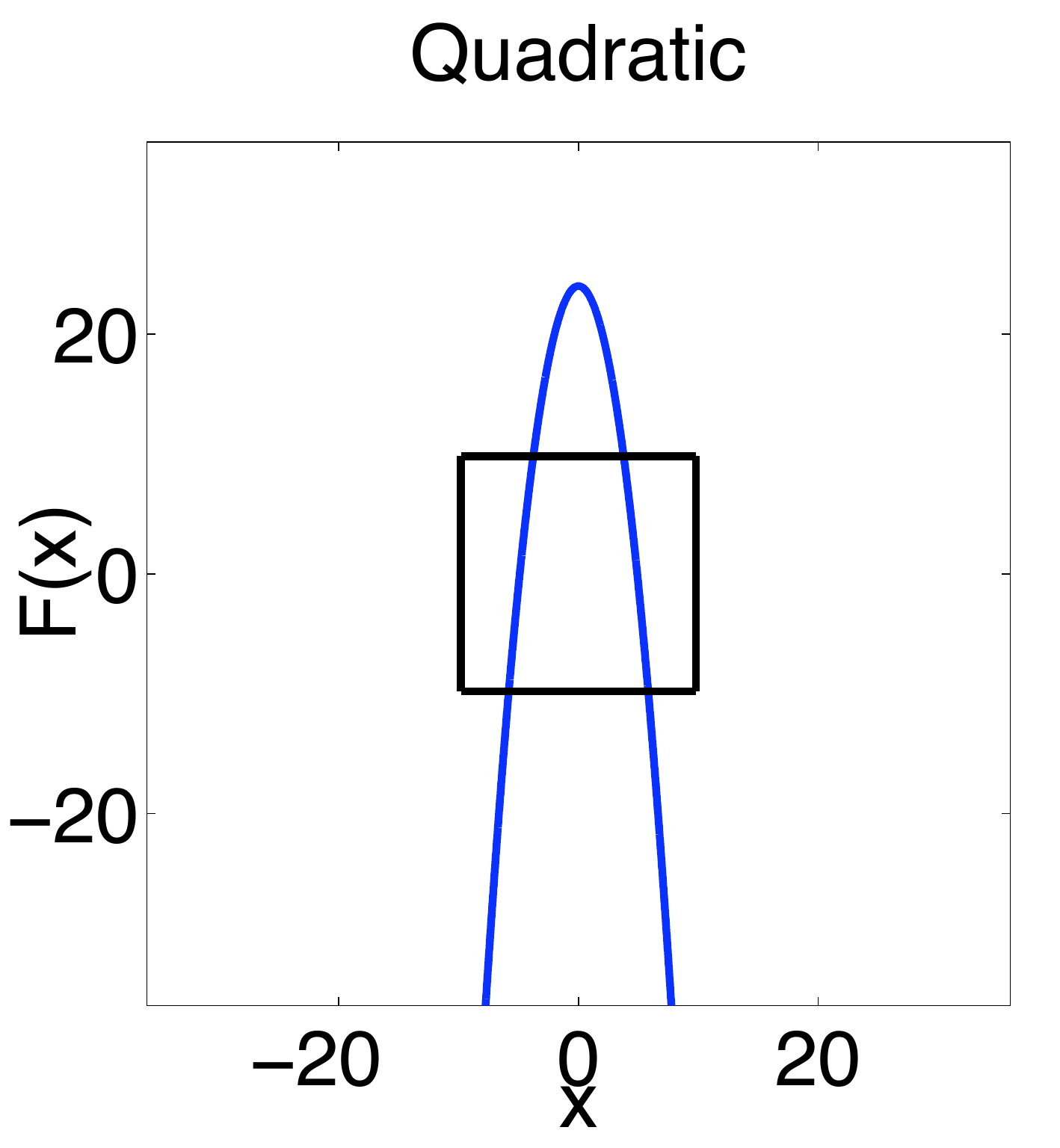}
\includegraphics[width=.24\textwidth]{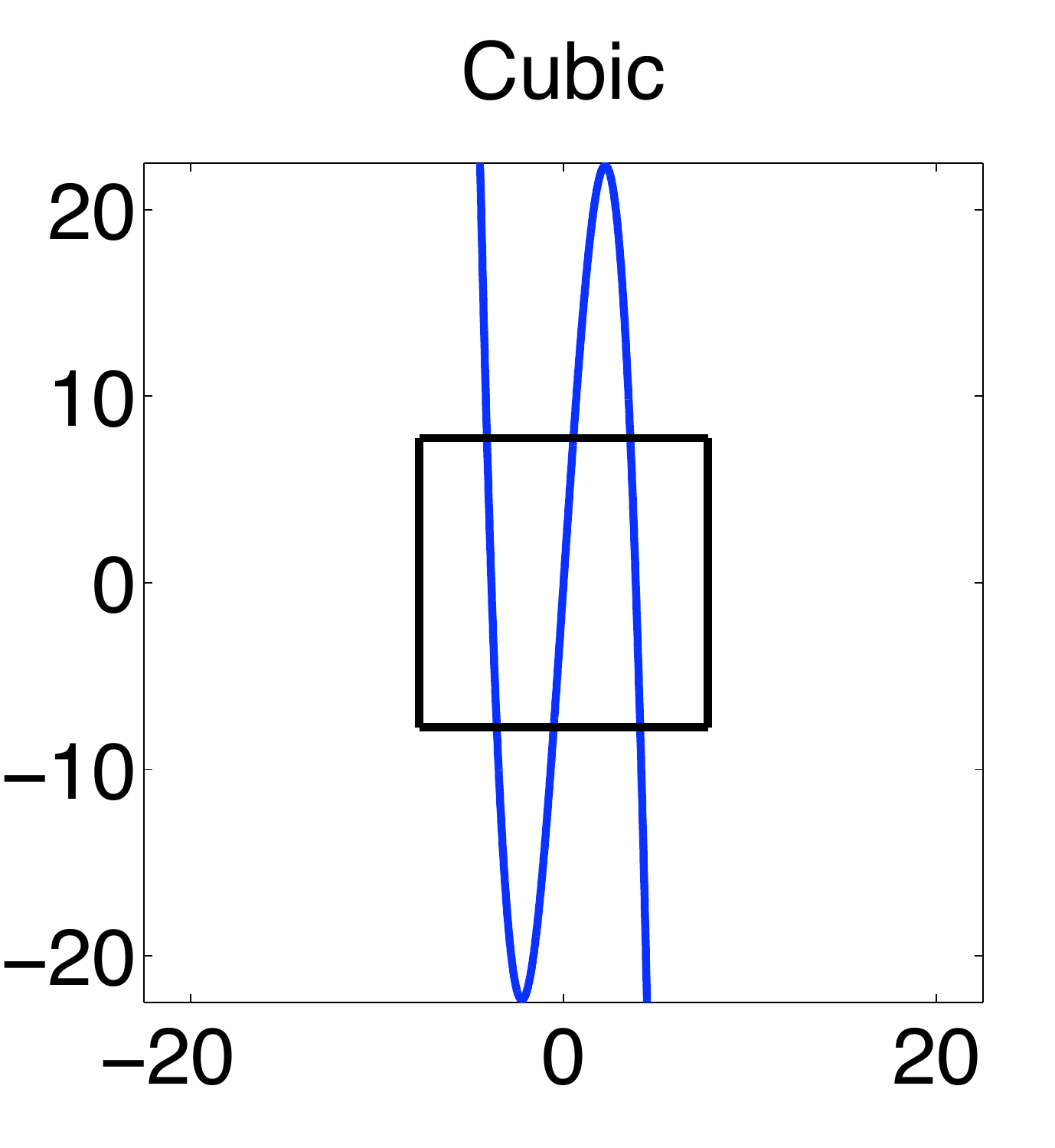}
\includegraphics[width=.24\textwidth]{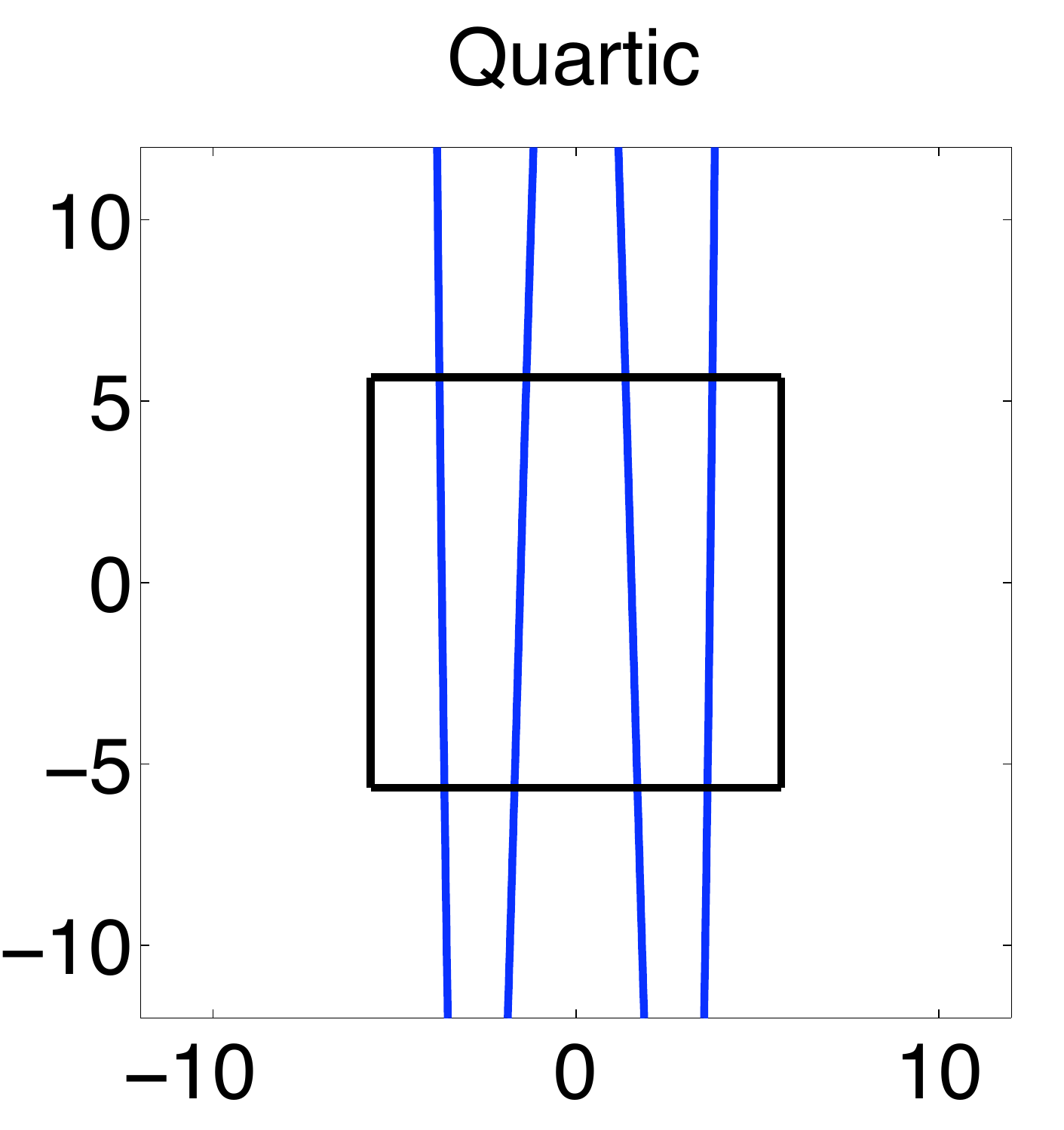}
\caption{\label{fig:threemaps} {\bf Quadratic, Cubic, and Quartic Maps.}
  The three maps from Eqn.~\ref{eqn:quad+g}, are shown for $g=0$ and for $\mu=24, 15,$ and $8$, respectively. The squares depicted are $[-2 \sqrt{\mu},2 \sqrt{\mu}]^2$. 
The maxima and minima 
are far larger than the size of the boxes, resulting in purely uniform 
chaotic behavior within the boxes. This is 
typical for large $\mu$. As $\mu$ increases, these three graphs would be 
stretched vertically. The values of the map $F$ for large $\mu$ at the critical points are proportional to $\mu$, 
$\mu^{3/2}$, and $\mu^2$, respectively, largely unaffected by $g$, which is bounded and has bounded derivative.
}
\end{center}
\end{figure}

Three rigorous examples of conservation of solitary cascades are
encapsulated in the following one-dimensional maps (see Figure~\ref{fig:threemaps}):
\begin{eqnarray}\label{eqn:quad+g} F(\mu,x) = \mu - x^2 +g(\mu,x)
&(\mbox{quadratic}),\\ F(\mu,x)= \mu x-x^3 + g(\mu,x)
&(\mbox{cubic}),\nonumber \\ F(\mu,x)=x^4- 2\mu x^2 + {\mu^2}/{2}+
g(\mu,x) &(\mbox{quartic}), \nonumber
\end{eqnarray} where $g$ is smooth, and for some real positive $\beta$,
\begin{eqnarray}\label{eqn:g1} |g(\mu,0)| < \beta &\mbox{ for all }
\mu, \mbox{ and } \\ |g_x(\mu,x)| < \beta &\mbox{ for all }
\mu,x. \nonumber
\end{eqnarray} 

It is straightforward to show that for $g \equiv 0$, each of these
maps has a $[\mu^*_1,\mu^*_2]$ such that there are no regular periodic
orbits for $F$ at  $\mu^*_1$, and $F$ has virtually uniform (PO) chaos at
$\mu^*_2$. This leads to the fact that for $g$ of the form given, each
of the three maps has no regular periodic orbits for $\mu_1$
sufficiently negative, and for $\mu_2$ sufficiently large has a
one-dimensional horseshoe map.  The conditions on $g$ guarantee that it does not
significantly affect the periodic orbits when $|\mu|$ is sufficiently
large; in particular it does not affect their eigenvalues, so it does
not affect the number of period-$p$ regular periodic orbits for large
$|\mu|$.  Hence  all three maps have no regular
periodic orbits for $\mu$ small and have no attracting
periodic orbits for $\mu$ large, and for sufficiently large
$|\mu|$, all periodic orbits are contained in the set $[-2
\sqrt{\mu},2 \sqrt{\mu}]$.  This leads to the following result:

\begin{theorem}[Conservation of cascades]  \label{theorem:conservation}
  For each of the functions $F$ in Eqn.~\ref{eqn:quad+g}, $F$ is
  generic for a residual set of $g$ chosen as in
  Eqn.~\ref{eqn:g1}. For each of the three examples, the number of
  stem-period-$k$ solitary cascades for all generic $F$ is independent
  of the choice of $g$.
\end{theorem}

In other words, fix $F$ to be one of the three types in
Eqn.~\ref{eqn:quad+g}.  Fix any $g$ as in Eqn.~\ref{eqn:g1} such that
$F$ is generic. Then there exist $\mu_L$ and $\mu_M$ such that as long
as $\mu_1<\mu_L$ and $\mu_2> \mu_M$, $F$ has the same number of
solitary cascades on $[\mu_1,\mu_2]$ as occur in the $g \equiv 0$ case
for the interval $[\mu^*_1,\mu^*_2]$.

As an interpretation of this statement take a set $B$ as large as you
like in parameter cross phase space, and make $g=-(\mu-x^2)$ in $B$,
so $F \equiv 0$ in $B$. Hence the only periodic orbit in $B$ is the
fixed point $x=0$, so $B$ contains no cascades. It might seem that we
have annihilated all cascades by this process, but the theorem
guarantees that every single one of these solitary cascades will
appear. They are just displaced from their original location, moved
outside of $B$.

The conservation principle works for higher-dimensional maps as
well. For example, we have shown in~\cite{sander:yorke:09b} that
even large-scale perturbations of the two-dimensional H{\'e}non map have
conservation of solitary cascades. That is, writing $x = (x_1, x_2)$,
\[
F(\mu,x_1,x_2) =
\left(
\begin{array}{c}
\mu +\beta x_2 -x_1^2+g(\mu,x_1, x_2)\\
x_1+h(\mu,x_1,x_2)
\end{array}
\right), \hspace{2.5cm} \mbox{(H\'enon)}
\] 
where $\beta$ is fixed, and the added function $(g, h)$ is smooth and
is very small for $||(\mu,x_1, x_2)||$ sufficiently large. See \cite{sander:yorke:09b} for a precise (and technical) formulation. Here
the added terms cannot destroy the stems which are unchanged in the
domain where $||(\mu, x_1, x_2)||$ is very large. Each stem must still
lead to its own solitary cascade.

If $F(\mu,x_1,\dots,x_N)$ is a set of $N$ coupled quadratic maps such
that
\[ x_i \mapsto K(\mu)-x_i^2+g(x_1,\dots,x_N), \hspace{2.5cm} 
\mbox{(coupled)} \] where $g$ is bounded
with bounded first derivatives, and $\lim_{k \to \pm
  \infty}K_i(\mu)=\pm \infty$, then there is conservation of
solitary stem-period-$k$ cascades.

\section{Off-on-off chaos for paired cascades}\label{sec:paired}

In the previous section, we concentrated on the implications of
Theorem~\ref{theorem:solitarypaired} on solitary cascades, but it also
has implications for paired cascades.
It describes a situation which is quite common in physical systems,
namely the case in which parameter regions with and without chaos are
interspersed, such as show in Figs.~\ref{Fig:vdp} and~\ref{Fig:pend}.
Specifically, we give the following definition describing the
situation where is no chaos at $\mu_1$ and $\mu_3$, while
at $F$ is chaotic at $\mu_2$. 

\begin{definition}[Off-on-off chaos]
Assume $F$ satisfies $A_1$--$A_4$ on $[\mu_1,\mu_2]$ and on
$[\mu_3,\mu_2]$, where $\mu_1<\mu_2<\mu_3$. Then $F$ is said to have
 {\bf off-on-off chaos} for $\mu_1<\mu_2<\mu_3$. 

\end{definition}

If $F$ has off-on-off chaos, we can apply 
Theorem~\ref{theorem:version2} to both $[\mu_1,\mu_2]$ and to $[\mu_2,\mu_3]$ and conclude
that there are infinitely many cascades in each of the two
intervals. The following theorem is stronger, in that we also conclude
that virtually all regular periodic points at  $\mu_2$ are contained
in paired cascades.
 
\begin{theorem}[Off-On-Off Chaos] \label{theorem:offonoff}
If $F$ has off-on-off chaos on $\mu_1<\mu_2<\mu_3$, then $F$ has infinitely many
(bounded) paired cascades and at most finitely many solitary cascades in $(\mu_1,\mu_3)$.
\end{theorem}

The result follows directly from combining the results of
Theorems~\ref{theorem:version2} and~\ref{theorem:solitarypaired}.

\begin{example}
Our numerical studies indicate that the time-$2\pi$ maps of the forced
double-well Duffing (Fig.~\ref{Fig:vdp}) and forced damped pendulum
(Fig.~\ref{Fig:pend}) have off-on-off chaos, and that it occurs on
multiple non-overlapping parameter regions. Here $2\pi$ is the forcing
period. We cannot prove there are only finitely many periodic
attractors, though we find very few. These systems have no periodic
repellers.
\end{example}

\section{Discussion}\label{sec:discussion}

\bigskip
\noindent
{\bf Our results in the context of routes to chaos.}
We now contrast our view with the traditional view of many different
routes to chaos.  People write about Routes to Chaos - where chaos
does indeed mean there is a chaotic attractor. Below we list classes
of distinct routes to chaotic attractors. Our results say that for
generic smooth maps depending on a parameter, there is a unique route
from no chaos to chaos (which we have defined to mean virtually uniform
PO chaos) -- in the sense of having a chaotic set that need not be
attracting, for example when there is a transverse homoclinic
point. Between a parameter value where there is no chaos and a
parameter value where there is chaos, there must be infinitely many
period-doubling cascades.

\pagebreak
\noindent {\bf Routes to a chaotic attractor} 
\begin{enumerate}
\item A chaotic attractor develops where there was no previous
  horseshoe dynamics; this includes what we might call the Feigenbaum
  cascade route.
\item There is a transient chaos set and a simple non-chaotic
  attractor (equilibrium point or periodic orbit) and that attractor
  becomes unstable. For periodic orbits there are three ways to become
  unstable.
\item Like above except the attractor is a torus with quasiperiodic
  dynamics. How many ways can a torus become unstable? We suspect in many ways.
\item Crisis route: as a parameter decreases, a chaotic
  attractor collides with its basin boundary and the chaotic set is no
  longer attracting. How many ways can this happen? Also unknown.
\item There is a chaotic attractor and a non-chaotic attractor and as
  a parameter is varied, the initial condition being used migrates
  into the basin of the chaotic attractor.
\item   Homoclinic explosions lead to chaotic dynamics.
\end{enumerate}

However, using the viewpoint described in this paper, 
{\bf there is only one route to chaos}. 

\bigskip
\noindent
{\bf Relating PO chaos and positive entropy.}  
Consider the following definition, related to exponential growth of
periodic orbits, alluded to when the definition was given: If
$|fixed(p)| \sim G^p$ for large $p$, then we say $G$ is the growth
factor, or we call $\log G$ the {\bf periodic orbit
  entropy}. Taking logs of both sides, dividing by $p$, and taking
limits, we get
\[ \mbox{limsup}_{p \to \infty}  \frac{\log |fixed(p)|}{p} = \log G
= h.\] We believe that many of the methods used here will lead to a
fruitful study of periodic orbit entropy.  Note that PO chaos occurs
exactly when $h > 0$. If there are only finitely many orbits for
example, $h= 0$. If there is at most a fixed number $k$ of periodic
orbits of each period $p$, again $h = 0$, even if there may be
infinitely many orbits. In many situations, there is a relationship
between exponential periodic growth and positive topological entropy,
and in fact in some cases periodic orbit entropy is equal to
topological entropy. We elaborate below.

Bowen showed~\cite{bowen:70} that for Axiom A diffeomorphisms, you can
find topological entropy exactly by looking at the growth rate for the
number of period-$p$ orbits as $p$ goes to infinity.  For H{\'e}non-like
maps, Wang and Young proved that the topological entropy coincides
with the exponential growth rate of the number of periodic points of
period $p$~\cite{wang:young:01}.  Chung and
Hirayama~\cite{chung:hirayama:03} show that for any surface
diffeomorphism with H{\"o}lder continuous derivative, the topological
entropy is equal to the exponential growth rate of the number of
hyperbolic periodic points of saddle type. That is, you can throw away
the attractors and repellers.

In terms of interval maps: Misiurewicz and
Szlenk~\cite{misiurewicz:szlenk:80} proved that if $f$ is a continuous
and piecewise monotone map of the interval, then the topological
entropy is bounded above by the exponential growth rate of the number
of periodic orbits. Building on~\cite{katok:mezhirov:98},
Chung~\cite{chung:01} showed that if $f$ is $C^{1+\alpha}$, similar
results can be obtained if one counts only the set of hyperbolic
periodic orbits (namely, periodic points $x$ for which $\vert (f^
p)'(x)\vert^{1/p} > 1$) or the set of transversal homoclinic points of
a source.

\bigskip
\noindent
{\bf Non-smooth maps.}
Continuous maps $F$ which are piecewise smooth
but not smooth violate our assumptions, but such maps can be thought
of as the limits of generic smooth maps $F_n$ which differ from $F$ only very near the
discontinuities of $F_x$. In dimension one, $F = \mu + b x + c |x|$
gives a rich collection of examples obtained by choosing constants $b$
and $c$ in interesting ways. See the extensive literature on border
collision bifurcations~\cite{nusse:ott:yorke:94}. Such maps deserve more discussion than we can give
here but we mention two examples. 

(1) For the tent map $F(\mu,x) = \mu - 2 |x|$, there are no periodic
orbits when $\mu < 0$; there is one periodic orbit, a fixed point at
0, when $\mu = 0$, and orbits of all periods when $\mu > 0$. All
orbits are in families of straight line rays that bifurcate from and
originate at $(0,0)$, the map's only bifurcation periodic orbit. In
this case, all bifurcations in all the cascades that would exist for
the approximating generic families $F_n$ tend to $(0,0)$ as $n \to
\infty$.

(2) For the tent map with slope $\pm\mu$, namely $F(\mu,x) = 1 -\mu
|x|$ for $1 < \mu \le 2$, a periodic orbit suddenly appears at $x=0$
for countably many $\mu$. For example if $\mu_3$ denotes the smallest
parameter with a period-three orbit, then the point $x=0$ is a period-three point, and only infinitely many periodic orbits whose periods are multiples of three bifurcate from it. All of the cascades for those
orbits that we would expect for a smooth map are collapsed into the
single point $(\mu_3, x=0)$.

\section*{Acknowledgments}
%
%
We thank Safa Motesharrei for his corrections and detailed comments. 
E.S.~was partially supported by NSF Grants DMS-0639300
and DMS-0907818, as well as NIH Grant R01-MH79502. 
J.A.Y.~was partially supported by NSF Grant DMS-0616585 and NIH Grant
R01-HG0294501.

%
%

\addcontentsline{toc}{section}{References}
\footnotesize
%
%

%
\end{document}